\documentclass[prd,showpacs,superscriptaddress,nofootinbib,floatfix,12pt]{revtex4-1} 
\usepackage[utf8]{inputenc}
\usepackage{amsmath,amssymb}
\usepackage{slashed}
\usepackage{graphicx}
\usepackage{subfigure}
\usepackage[colorlinks]{hyperref}
\usepackage[usenames,dvipsnames]{color}

\usepackage{mathrsfs}
\usepackage{color}
\usepackage{bbm}               
\usepackage{xspace}				
\usepackage{epstopdf}
\usepackage{pgffor}		
\usepackage{array}
\usepackage{multirow}

\begin{document}
\title{Mass Spectra of Full-Heavy and Double-Heavy Tetraquark States in the Conventional Quark Model}

\author{Qi Meng}\email{qimeng@nju.edu.cn}
\affiliation{Department of Physics, Nanjing University, Nanjing 210093, P.R. China}

\author{Guang-Juan Wang}\email{wgj@post.kek.jp}
\affiliation{KEK Theory Center, Institute of Particle and Nuclear Studies (IPNS), High Energy Accelerator Research Organization (KEK), 1-1 Oho, Tsukuba, Ibaraki, 305-0801, Japan}

\author{Makoto Oka}\email{makoto.oka@riken.jp}
\affiliation{Nishina Center
for Accelerator-Based Science, RIKEN, Wako 351-0198, Japan}
\affiliation{Advanced Science Research Center, Japan Atomic Energy
Agency, Tokai, Ibaraki, 319-1195, Japan}

\begin{abstract}

A comprehensive study of the $S$-wave heavy tetraquark states with identical quarks and antiquarks, specifically $QQ\bar Q'\bar Q'$ ($Q, Q'=c,b$),  $QQ\bar s\bar s$/$\bar Q\bar Q ss$, and $QQ\bar q\bar q$/$\bar Q\bar Q qq$ ($q=u,d$), are studied in a unified constituent quark model. This model contains the one-gluon exchange and confinement potentials. The latter is modeled as the sum of all two-body linear potentials. We employ the Gaussian expansion method to solve the full four-body Schr{\"{o}}dinger equations, and search bound and resonant states using the complex-scaling method. We then identify $3$ bound and $62$ resonant states. The bound states are all $QQ\bar q\bar q$ states with the isospin and spin-parity quantum numbers $I(J^P)=0(1^+)$: two bound $bb\bar{q}\bar{q}$ states with the binding energies, $153$ MeV and $4$ MeV below the $BB^*$ threshold, and a shallow $cc\bar{q}\bar{q}$ state at $-15$ MeV from the $DD^*$ threshold. The deeper $bb\bar q \bar q$ bound state aligns with the lattice QCD predictions, while $cc\bar q\bar q$ bound state, still has a much larger binding energy than the recently observed $T^+_{cc}$ by LHCb collaboration. No bound states are identified for the $QQ\bar Q'\bar Q'$, $QQ\bar s\bar s$ and $QQ\bar q\bar q$ with $I=1$. Our analysis shows that the bound $QQ\bar{Q}'\bar{Q}'$ states are more probable with a larger mass ratio, $m_Q/m_{Q'}$. Experimental investigation for these states is desired, which will enrich our understanding of hadron spectroscopy and probe insights into the confinement mechanisms within tetraquarks.

\end{abstract}
\maketitle

\section{introduction}

In the past decades, numerous exotic hadrons that challenge the conventional picture of quark-antiquark meson and three-quark baryon structure have been observed by various collaborations such as BABAR, Belle, BES, LHCb, CMS, ATLAS, etc. These exotic hadrons are good candidates of the multiquark states, such as tetraquarks ($qq\bar q \bar q$) and pentaquarks ($qqqq\bar q $). Unlike the conventional mesons and baryons, the multiquark states possess novel color configurations and exhibit a poorly understood confinement mechanism. 
It then obscures the interpretations of the exotic states that may originate from hadronic molecules or compact tetraquarks.
Extensive research has been conducted to unravel the nature of these exotic states.  For a comprehensive overview, we refer the readers to recent reviews \cite{Chen:2016qju,Chen:2016spr,Lebed:2016hpi,Esposito:2016noz,Hosaka:2016pey,Guo:2017jvc,Ali:2017jda,Liu:2019zoy,Brambilla:2019esw,Lucha:2021mwx,Chen:2021ftn,Chen:2022asf,Meng:2022ozq}. 

The discovery of the full-heavy tetraquark state $QQ\bar Q'\bar Q'$ ($Q, Q'=c,b$) has cast new light on the mysterious confinement mechanism. They are good candidates for the compact tetraquark state, since the interactions among the heavy quarks are likely dominated by the short-range one-gluon-exchange rather than the long-range potentials by exchanging light meson. Significant observations have been made in the $cc\bar c\bar c$ states. The LHCb collaboration reported a broad structure in the 6.2$\sim$6.8 GeV range and a narrow resonance, $X(6900)$, in the di-$J/\psi$ invariant mass spectrum~\cite{LHCb:2020bwg}. The $X(6900)$ was subsequently confirmed by both the CMS \cite{CMS:2023owd} and ATLAS \cite{ATLAS:2023bft} collaborations. Furthermore, the CMS identified two new resonances $X(6600)$ and $X(7200)$, while ATLAS reported $X(6400)$, $X(6600)$ and $X(7200)$ \cite{ATLAS:2023bft}. 
The full-bottom tetraquark, on the other hand, still lacks experimental observation. Both the LHCb~\cite{Aaij:2018zrb} and CMS~\cite{Khachatryan:2016ydm,Durgut} collaborations have searched the full-bottom tetraquark in the $\Upsilon(1S)\mu^+\mu^-$ invariant mass spectrum but did not observe a significant excess for a $bb\bar b \bar b$ state.

The theoretical investigations on the above full-heavy quark state $QQ\bar Q'\bar Q'$ within the tetraquark framework have started decades ago~\cite{Iwasaki:1975pv,Chao:1980dv,Ader:1981db,Zouzou:1986qh,Heller:1986bt,SilvestreBrac:1992mv,SilvestreBrac:1993ry}, employing various methodologies, for instance, the diffusion Monte Carlo approach \cite{Bai:2016int,Gordillo:2020sgc,Ma:2022vqf,Gordillo:2020sgc}, the QCD sum rules \cite{Zhang:2020xtb,Yu:2022lak,Wu:2022qwd},  the effective field theory \cite{Zhang:2020hoh,Caswell:1985ui,Bodwin:1994jh}, the  Bethe-Salpeter equations \cite{Li:2021ygk,Ke:2021iyh,Chen:2022sbf}, the different quark models \cite{Tiwari:2021tmz,Scarpa:2019fol,An:2022fvs,Anwar:2023svj,Wang:2023jqs,Wang:2022yes,Faustov:2022mvs}, and the other phenomenological models. Among these methods, the constituent quark model stands out for its success and effectiveness in investigating the spectra of the mesons and the baryons \cite{Godfrey:1985xj,Capstick:1985xss}.  Notice that, besides the tetraquark model, other interpretations for the observed $cc\bar c\bar c$ states also exist, such as the dynamical rescattering mechanism of double-charmonium \cite{Dong:2020nwy,Guo:2020pvt,Gong:2020bmg,Wang:2020wrp,Kuang:2023vac}. For more details, we refer to Ref. \cite{Chen:2022asf} for a review.

For the double-heavy tetraquark state $QQ\bar q\bar q$, the LHCb collaboration observed a narrow $T^+_{cc}$ state which is located remarkably close to the $DD^*$ threshold \cite{LHCb:2021vvq,LHCb:2021auc}. This state has been widely investigated, often being interpreted as a molecular state ~\cite{Agaev:2021vur,Ling:2021bir,Chen:2021vhg,Dong:2021bvy,Feijoo:2021ppq,Yan:2021wdl,Xin:2021wcr,Huang:2021urd,Fleming:2021wmk,Azizi:2021aib,Hu:2021gdg,Chen:2021cfl,Albaladejo:2021vln,Du:2021zzh,Deng:2021gnb,Agaev:2022ast,Braaten:2022elw,He:2022rta,Abreu:2022lfy,Achasov:2022onn,Mikhasenko:2022rrl,Wang:2022jop,Lyu:2023xro}.  In the bottom sector, the experimental data is still absent. However, all the lattice QCD calculations predicted the existence of the tetraquark states $bb\bar q\bar q$ ($q=u,d$) featuring substantial binding energies ranging from about $-60$ to $-190$ MeV~\cite{Bicudo:2015kna,Bicudo:2016ooe,Aoki:2023nzp,Francis:2016hui,Bicudo:2022cqi}. Additionally, numerous theoretical studies have been conducted on other $bb\bar q\bar q$ states within both the molecular \cite{Ren:2023pip,Ohkoda:2012hv,Li:2012ss,Manohar:1992nd,Bicudo:2016ooe} and the compact tetraquark models \cite{Meng:2023for,Karliner:2013dqa,Eichten:2017ffp,Yang:2009zzp,Vijande:2009kj}, predicting the existence of the deeply bound $bb\bar q\bar q$ ($q=u,d$) state.  More details are referred to Ref. \cite{Chen:2022asf}. Such significant binding energies present a challenge to arrange them as molecular states, which are typically found near the threshold. Therefore, the study considers them as compact tetraquark states will help to explore the inner structures.

Although numerous quark model studies have been conducted on the tetraquark states, particularly focusing on the fully heavy tetraquark $QQ\bar Q'\bar Q'$ and double-heavy quark states $QQ\bar q\bar q$, inspired by the experimental findings and lattice QCD calculations~\cite{Tiwari:2021tmz,Scarpa:2019fol,Faustov:2020qfm,An:2022fvs,Anwar:2023svj, Wang:2023jqs,Wang:2022yes}. A comprehensive and systematic understanding of the tetraquark system still benefits from a unified quark model approach. In this paper, we systematically calculated the mass spectra of the S-wave fully tetraquark states $bb\bar{b}\bar{b}$,  $cc\bar{c}\bar{c}$ and $bb\bar{c}\bar{c}$ as well as the double-heavy tetraquark states $bb\bar q\bar q$, $cc\bar q\bar q$, $bb\bar s\bar s$, and $cc\bar s\bar s$, using a non-relativistic quark model and the Gaussian expansion method. These states contain identical quarks and antiquarks, which have much simpler configurations constrained by the Fermi-Dirac statistics compared to other systems. Additionally, we explore the dependence of bound state formation on the ratio of quark mass to antiquark mass.

In the prior quark model studies,  wave function expansion methods, particularly the Gaussian expansion method, have been extensively employed. In this approach, the tetraquark wave functions are expanded by a finite number of basis functions, effectively creating a limited spatial volume. Consequently, all states, including the bound, resonant, and scattering states, are manifested as  ``discrete states", leading to the misidentification of ``fake" resonances. To identify the genuine resonant states from the scattering states, real-scaling \cite{Simons:1981gbz,Hiyama:2005cf,Meng:2019fan} and complex-scaling methods have been utilized extensively \cite{Myo:2014ypa,Myo:2020rni,Moiseyev:1998gjp}. Notably, many eigenstates initially identified as resonances in the wave function expansion method, disappear when subjected to the real- and complex-scaling methods. This indicates they are predominantly scattering states, rather than resonant states. In this work, we use the complex-scaling method to identify the compact tetraquark states from the scattering states.

This paper is organized as follows: In Sec. \ref{sec1:Formalism}, we introduce the Hamiltonian for the tetraquark system, and the inter-quark potentials are derived using the quark model. We also present the Gaussian expansion method employed for the four-body calculations and the complex-scaling method to identify the bound and resonant states. In  Sec. \ref{sec:rd}, we presented the calculated mass spectra of both the bound and resonant states, along with discussions. Additionally, we investigate the mass spectrum of the lowest-lying states, either the bound or resonant states, in relation to the mass ratio of the heavy quarks and antiquarks. Finally, a brief summary of our findings and conclusions is given in Sec. \ref{sec:summary}.

\section{Formalism} \label{sec1:Formalism}

\subsection{Hamiltonian} \label{sec1.1:Hamiltonian}

The Hamiltonian for the tetraquark system reads 
\begin{eqnarray}
H=\sum_{i=1}^4 \Big(m_i+ \frac{\boldsymbol{p}_i^2}{2m_i} \Big)-T_G + \sum_{i<j=1}^4 \sum_{a=1}^8 \frac{\lambda_i^a}{2}  \frac{\lambda_j^a}{2} \big(-\frac{3}{4}V_{ij}(\boldsymbol{r}_{ij})\big)  ,
\end{eqnarray}
where $m_i$ and $\boldsymbol{p}_i$ are the mass and momentum of the $i$-th quark, respectively. $T_G$ denotes the kinetic energy of the center of mass. $\lambda_i^a$ is the color SU(3) Gell-mann matrix for the $i$-th quark with color index $a$. The quark-quark potentials $V_{ij}$ are obtained by using the quark model proposed by Semay and Silvestre-Brac \cite{Silvestre-Brac:1996myf}, and are defined as follows,
\begin{eqnarray}
V_{i j}(\boldsymbol{r})= & -\frac{\kappa}{r}+\lambda r^p-\Lambda +\frac{2 \pi \kappa^{\prime}}{3 m_i m_j} \frac{\exp \left(-r^2 / r_0^2\right)}{\pi^{3 / 2} r_0^3} \boldsymbol{\sigma}_i \cdot \boldsymbol{\sigma}_j,
\end{eqnarray}
with
\begin{eqnarray}
    r_0\left(m_i, m_j\right)=A\left(\frac{2 m_i m_j}{m_i+m_j}\right)^{-B}.
\end{eqnarray}
where $\boldsymbol{\sigma}$ is the spin operator of the quark or antiquark. This potential consists of the color-electric (color-Coulomb) potential, the linear confinement potential, and a constant term associated with the color-magnetic spin-spin hyperfine interaction term. In this study, we employed the AL1 set of parameters, the values of which are shown in Table \ref{tab:para} {\footnote{Throughout this work, we will refer to the quark model using the AL1 set of parameters as ``AL1" for simplicity.}. }

\begin{center}
\renewcommand\arraystretch{1.0} 
\begin{table}[h]
\caption{The AL1 set of parameters used in the quark model for quark-quark interactions \cite{Silvestre-Brac:1996myf}. }
\label{tab:para}
\begin{tabular}{p{2.5cm}<{\centering}p{2.5cm}<{\centering}  |p{2.5cm}<{\centering} p{2.5cm}<{\centering}  }
\hline\hline
\multicolumn{4}{c}{AL1} \\
\hline
$m_{u,d}$(GeV)& 0.315   & $\kappa$ & 0.5069 \\
$m_{s}$(GeV)& 0.577     & $\kappa'$& 1.8609 \\
$m_{c}$(GeV)& 1.836     & $\lambda($GeV$^{p+1})$& 0.1653 \\
$m_{b}$(GeV)& 5.227     & $\Lambda$& 0.8321 \\
$p$& 1 & $B$&0.2204  \\
&  & $A($GeV$^{B-1})$&1.6553  \\
\hline\hline
\end{tabular}
\end{table}
\end{center}

The mass spectra of heavy quarkonia and heavy mesons, calculated using the AL1 potential, along with their comparisons with the experimental values \cite{Workman:2022ynf},  are listed in Table \ref{tab:hq} and \ref{tab:hm}, respectively in Appendix \ref{appendix-mm}. The tables demonstrate that the AL1 potential effectively describes both the $Q\bar Q'$ and $Q\bar q$ systems. Additionally, for comparison, the charmonium mass spectrum for the BGS potential, as proposed by Barnes et al. \cite{Barnes:2005pb}, is also included. The BGS more accurately reproduces the masses of both ground and excited charmonia, aligning closely with experimental results. This accuracy was the key factor in using the BGS potential in our previous study on the mass spectrum of fully-charmed tetraquark $cc\bar{c}\bar{c}$ \cite{Wang:2022yes}.  However, BGS cannot reproduce the masses of ground and excited states of other mesons with the same parameters. Therefore, we employ the AL1 model to provide a more comprehensive description of not only $cc\bar c\bar c$ but also other $QQ\bar Q'\bar Q'$, $QQ\bar s\bar s$, and $QQ\bar q\bar q$ systems.

\subsection{Methodology}

The four-body Schr\"{o}dinger equation is solved by the Gaussian expansion method. The total wave function of a tetraquark, $\Psi_{I,JM}$, with isospin $I$, total spin $J$ and its $z$-component $M$ is given by 
\begin{eqnarray}
\psi_{J M}&= & \sum_{C=a, b} \sum_\alpha A_{12} A_{34} \sum_\alpha \mathcal{B}_\alpha^{(C)} \xi_C^{(C)} \eta_I^{(C)} \notag \\
&& \times\left[\left[\phi_{n l}\left(\boldsymbol{r}_C\right) \otimes \phi_{N L}\left(\boldsymbol{R}_C\right) \otimes \phi_{\nu \lambda}\left(\boldsymbol{\rho}_C\right)\right]_{J_L} \otimes \chi_S^{(C)}\right]_{J M},
\end{eqnarray}
where $\xi_C^{(C)}$, $\eta_I^{(C)}$, and $\chi_S^{(C)}$ represent the color, isospin and spin wave functions, respectively.  The spatial wave functions in three independent Jacobi coordinates are expressed as $\phi_{n l}\left(\boldsymbol{r}_C\right)$, $\phi_{N L}\left(\boldsymbol{R}_C\right)$ and $\phi_{\nu \lambda}\left(\boldsymbol{\rho}_C\right)$.
The superscript $C$ denotes specific diquark-antidiquark (a) and meson-meson (b) configurations as illustrated in Fig. 1 in Ref. \cite{Wang:2022yes}. The $A_{12}$ and $A_{34}$ are the antisymmetrization operators for the identical quarks. $\mathcal{B}_\alpha^{(C)}$ is the expansion coefficient where the index $\alpha$ denotes all the basis functions characterized by the set of quantum numbers $\{ n,l, N, L,\nu,\lambda, J_L, S \}$. These quantum numbers can form the total $I(JM)$ quantum numbers.

In this study, we focus on the S-wave tetraquark states. The three orbital angular momentum along all the three Jacobi coordinates are assumed to be $0$. 
In the $QQ\bar{Q}'\bar{Q}'$, $QQ\bar s\bar s$, and $QQ\bar q\bar q$ systems, both the two quarks and antiquarks are identical, respectively.  Constrained by Fermi-Dirac statistics,
the possible flavor spin-color wave functions of the tetraquarks are given by
\begin{eqnarray}
\chi_{a,1}^{0^{+}}&=\Big[ \big\{ QQ \big\}_{\bar{3}_c}^{s=1} \big\{ \bar{Q}'\bar{Q}' \big\}_{3_c}^{s=1} \Big]^{S=0} , \\
\chi_{a,2}^{0^{+}}&=\Big[ \big\{ QQ \big\}_{6_c}^{s=0} \big\{ \bar{Q}'\bar{Q}' \big\}_{\bar{6}_c}^{s=0} \Big]^{S=0} , \\
\chi_{b,1}^{0^{+}}&=\Big[ \big\{ Q\bar{Q}' \big\}_{1_c}^{s=1} \big\{ Q\bar{Q}' \big\}_{1_c}^{s=1} \Big]^{S=0} , \\
\chi_{b,2}^{0^{+}}&=\Big[ \big\{ Q\bar{Q}' \big\}_{1_c}^{s=0} \big\{ Q\bar{Q}' \big\}_{1_c}^{s=0} \Big]^{S=0} , \\
\chi_{a,1}^{1^{+}}&=\Big[ \big\{ QQ \big\}_{\bar{3}_c}^{s=1} \big\{ \bar{Q}'\bar{Q}' \big\}_{3_c}^{s=1} \Big]^{S=1} , \\
\chi_{b,1}^{1^{+}}&=\Big[ \big\{ Q\bar{Q}' \big\}_{1_c}^{s=1} \big\{ Q\bar{Q}' \big\}_{1_c}^{s=0} \Big]^{S=1} , \\
\chi_{b,2}^{1^{+}}&=\Big[ \big\{ Q\bar{Q}' \big\}_{1_c}^{s=1} \big\{ Q\bar{Q}' \big\}_{1_c}^{s=1} \Big]^{S=1} , \\
\chi_{a,1}^{2^{+}}&=\Big[ \big\{ QQ \big\}_{\bar{3}_c}^{s=1} \big\{ \bar{Q}'\bar{Q}' \big\}_{3_c}^{s=1} \Big]^{S=2} , \\
\chi_{b,1}^{2^{+}}&=\Big[ \big\{ Q\bar{Q}' \big\}_{1_c}^{s=1} \big\{ Q\bar{Q}' \big\}_{1_c}^{s=1} \Big]^{S=2}.
\end{eqnarray}
In the  $QQ\bar Q'\bar Q'$, $\bar Q\bar Qss$, and $QQ\bar q\bar q$ with isospin $I=1$ systems, the color-flavor-spin configurations are identical, except for the $cc\bar c\bar c$ and $bb\bar b\bar b$. The possible $J^{PC}$ quantum numbers for $cc\bar c\bar c$ and $bb\bar b\bar b$ are $J^{PC}=0^{++}$, $1^{+-}$ and $2^{++}$. In the $1^{+-}$ states, the  $\chi_{b,2}^{1^{+}}$ is prohibited due to its $J^{PC}=1^{++}$.
The systems $bb\bar c\bar c$, $QQ\bar s\bar s$, and $I=1$ $QQ\bar q\bar q$ with different quark and antiquark, exhibit the possible $J^P=0^{+}$, $1^{+}$ and $2^{+}$. In the S-wave double-heavy tetraquark state $QQ\bar q\bar q$ with $I=0$, the only possible  $J^P$ quantum number is $1^+$, and its flavor spin-color wave functions are given by
\begin{eqnarray}
\chi_{a,1}^{1^{+}(I=0)}&=\Big[ \big\{ QQ \big\}_{\bar{3}_c}^{s=1} \big\{ \bar{q}\bar{q} \big\}_{3_c}^{s=0} \Big]^{S=1} , \\
\chi_{a,2}^{1^{+}(I=0)}&=\Big[ \big\{ QQ \big\}_{6_c}^{s=0} \big\{ \bar{q}\bar{q} \big\}_{\bar{6}_c}^{s=1} \Big]^{S=1} , \\
\chi_{b,1}^{1^{+}(I=0)}&=\Big[ \big\{ Q\bar{q} \big\}_{1_c}^{s=1} \big\{ Q\bar{q} \big\}_{1_c}^{s=0} \Big]^{S=1} , \\
\chi_{b,2}^{1^{+}(I=0)}&=\Big[ \big\{ Q\bar{q} \big\}_{1_c}^{s=1} \big\{ Q\bar{q} \big\}_{1_c}^{s=1} \Big]^{S=1}.
\end{eqnarray}

The complex-scaling method (CSM) is applied to identify the bound and resonant states separately from the scattering states ~\cite{Myo:2014ypa,
Myo:2020rni,Aoyama:2006,moiseyev1998quantum,Myo:2014ypa}. In CSM, the relative coordinates $\boldsymbol{r}$ between quarks and the corresponding relative momentum $\boldsymbol{p}$ are transformed into the complex-scaled forms,
\begin{eqnarray}
\boldsymbol{r}\rightarrow \boldsymbol{r}e^{i\theta}, \\ \boldsymbol{p}\rightarrow \boldsymbol{p}e^{-i\theta},
\end{eqnarray}
where  $\theta$ represents a positive constant angle.
Consequently, both the Hamiltonian and the corresponding Schr\"{o}dinger equation are  complex-scaled and  are expressed as follows:

\begin{eqnarray}
\begin{aligned}
H_\theta & =H\left(\boldsymbol{r} e^{i \theta}, \boldsymbol{p} e^{-i \theta}\right) \\
& =\sum_{i=1}^4 \frac{\boldsymbol{p}_i^2}{2 m_i} e^{-i 2 \theta}+\sum_i m_i+\sum_{i<j=1}^4 \frac{\lambda_i}{2} \frac{\lambda_j}{2} V_{i j}\left(\boldsymbol{r}_{i j} e^{i \theta}\right).
\end{aligned}
\end{eqnarray}
and
\begin{eqnarray}
H_{\theta}\Psi_{\theta}=E_{\theta}\Psi_{\theta}.
\end{eqnarray}
 The eigenvalues of ground, resonance, and scattering states are discrete and complex. According to the ABC theorem \cite{Aguilar:1971ve,Balslev:1971vb}, as the $\theta$ varying, eigenvalues of the scattering states rotate with $2\theta$ along the scattering line. In contrast, the eigenvalues of bound states and resonances remain stable and do not move with the changes in $\theta$. On the complex energy plane, the bound states appear on the real axis below the lowest threshold. When the condition $\theta \geq \frac{1}{2} {\rm{tan}}^{-1}(\Gamma/2 E_r)$ is satisfied, the resonances can be identified with their complex eigenvalues $E=E_{r}-i\Gamma/2$ where $E_r$ is the resonance energy and $\Gamma$ is the decay width.

\section{Results and discussions} \label{sec:rd}

\begin{figure}[h]
  \centering
  \includegraphics[width=14cm]{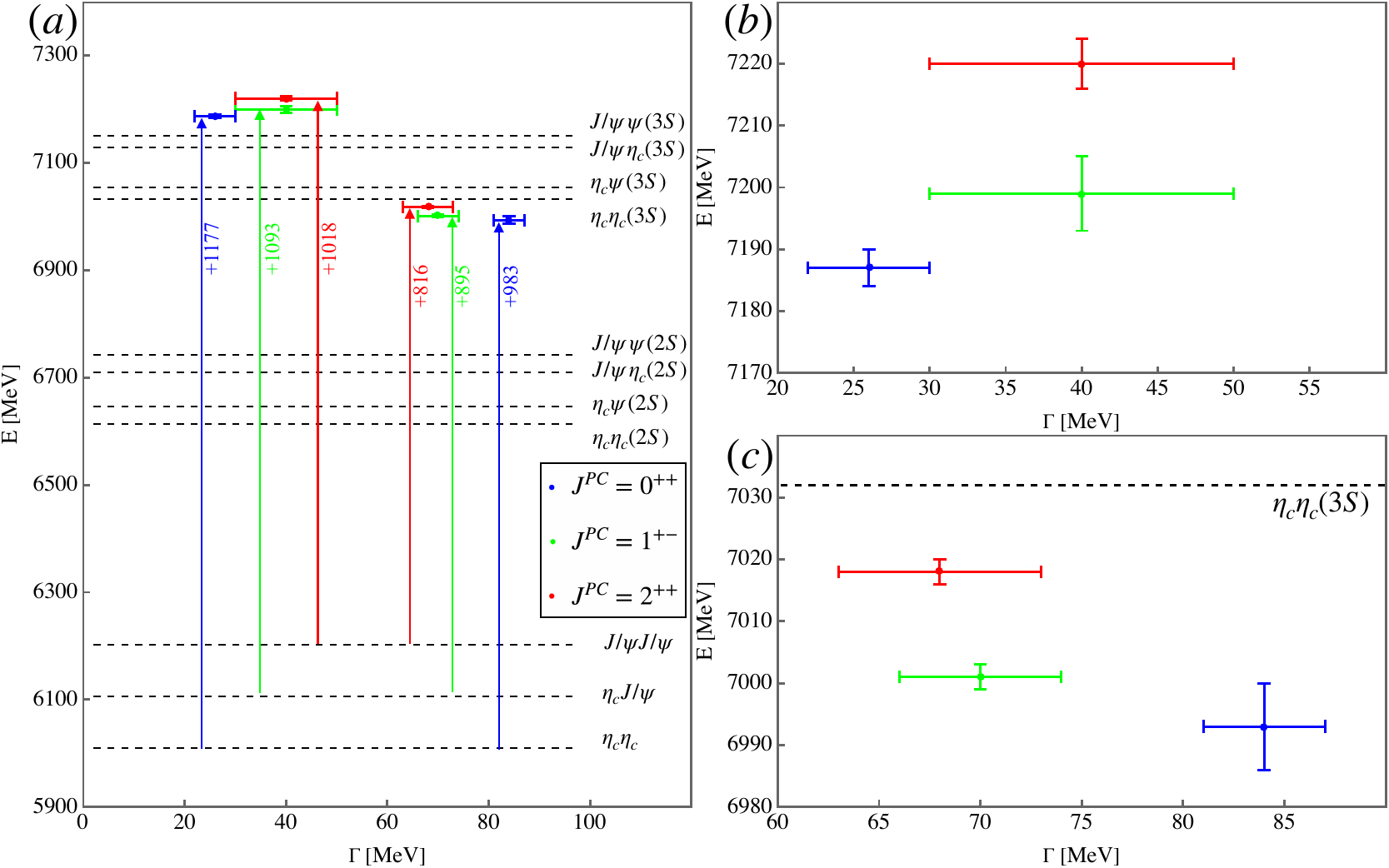}
    \caption{ The masses and widths of the $cc\bar{c}\bar{c}$ resonances obtained in the complex-scaling method and their positions relative to scattering states are shown in Subfigure $(a)$. Subfigures $(b)$ and $(c)$  provide a detailed view of the areas around the resonant pole positions to clarify uncertainties. The central values of the resonances are determined by averaging the results obtained with four different scaling constant $\theta$ values in CSM, and the uncertainties represent the variation from these central values.}
  \label{ELcccc}
\end{figure}

\begin{figure}[h]
  \centering
  \includegraphics[width=14cm]{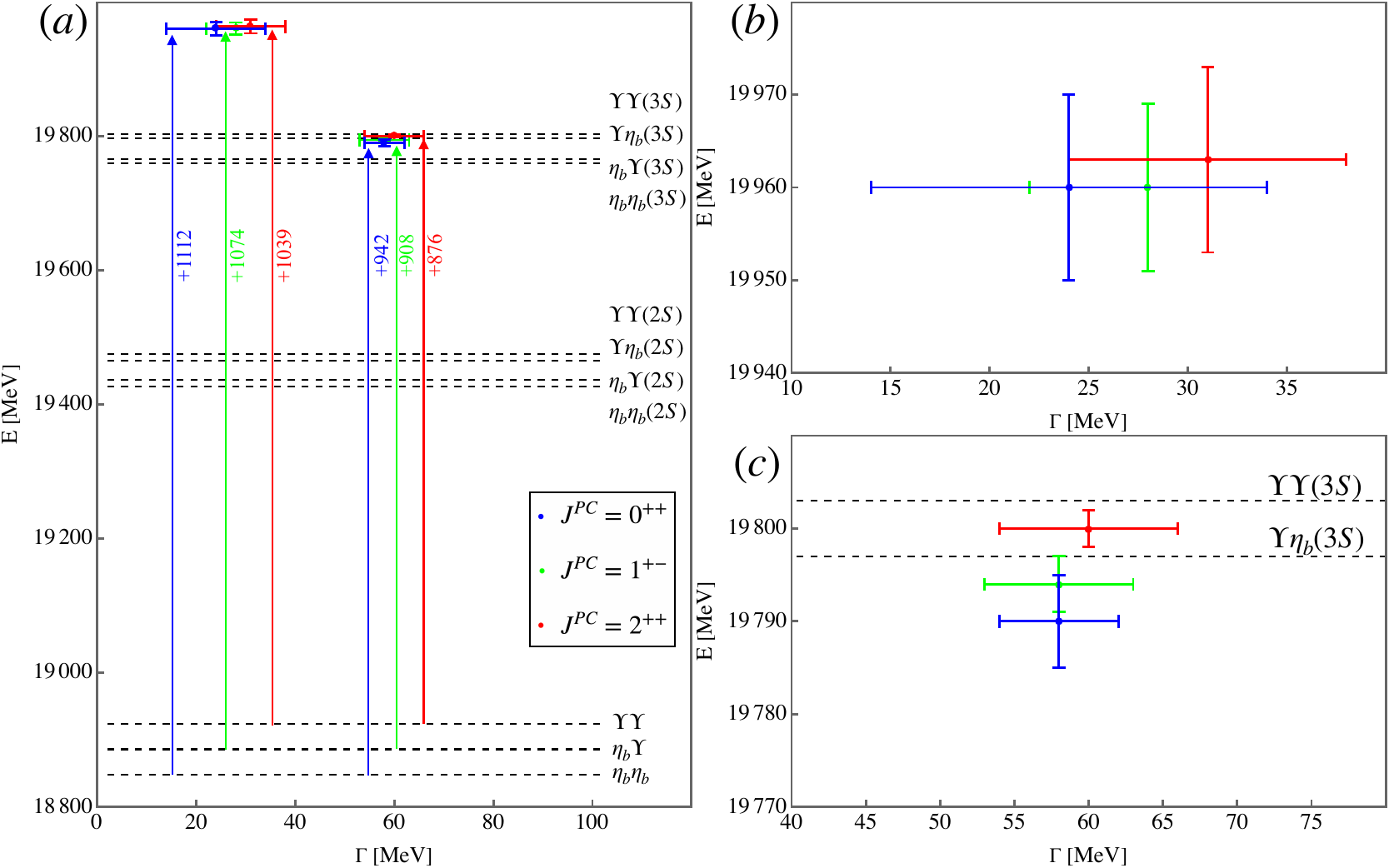}
    \caption{ The masses and widths of the $bb\bar{b}\bar{b}$ resonances obtained in the complex-scaling method and their positions relative to scattering states. For more details, see  caption of Fig. \ref{ELcccc}.}
  \label{ELbbbb}
\end{figure}

\begin{figure}[h]
  \centering
  \includegraphics[width=14cm]{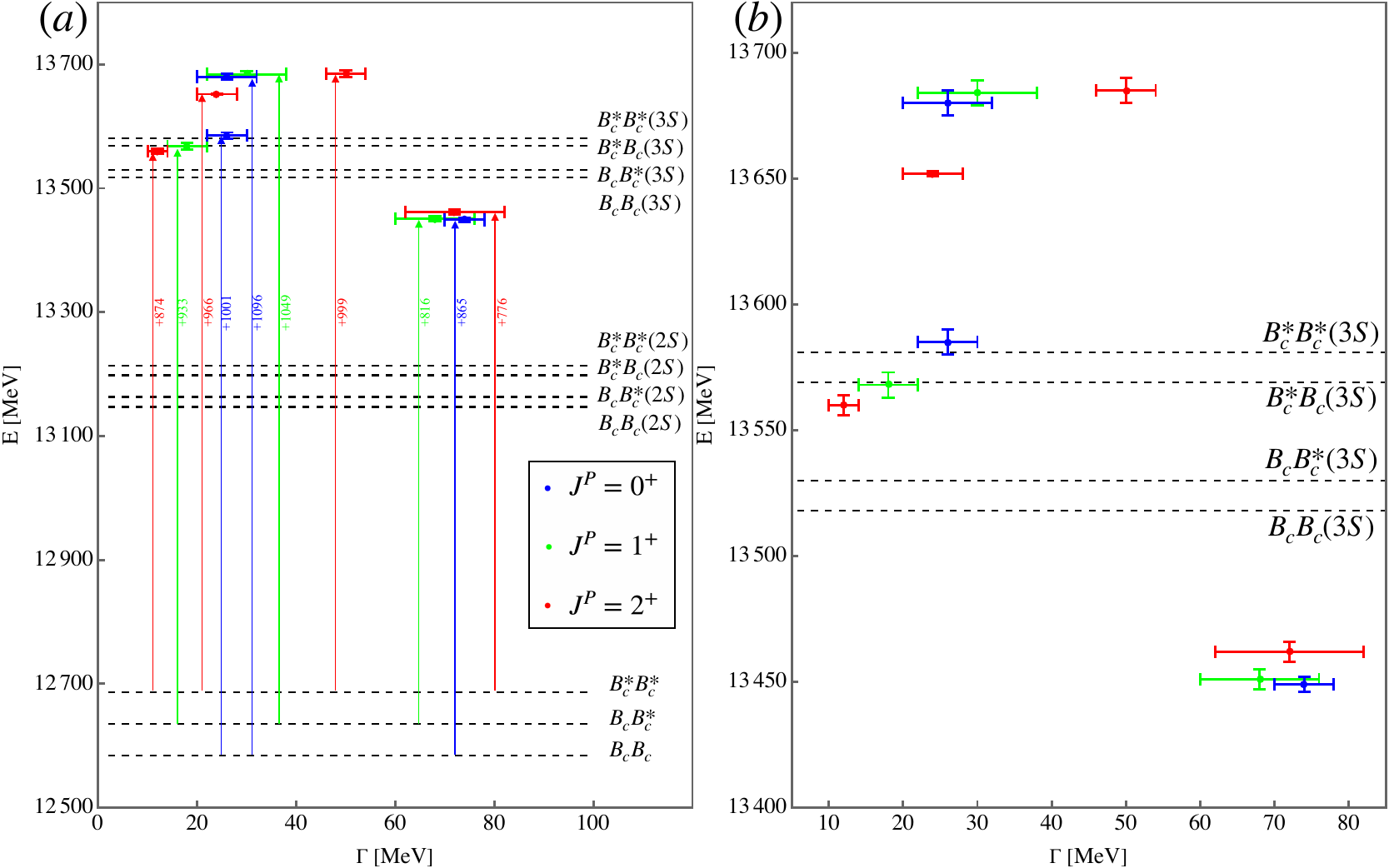}
    \caption{ The masses and widths of the $bb\bar{c}\bar{c}$ resonances obtained in the complex-scaling method and their positions relative to scattering states. For more details, see  caption of Fig. \ref{ELcccc}.}
  \label{ELbbcc}
\end{figure}

\begin{figure}[h]
  \centering
  \includegraphics[width=14cm]{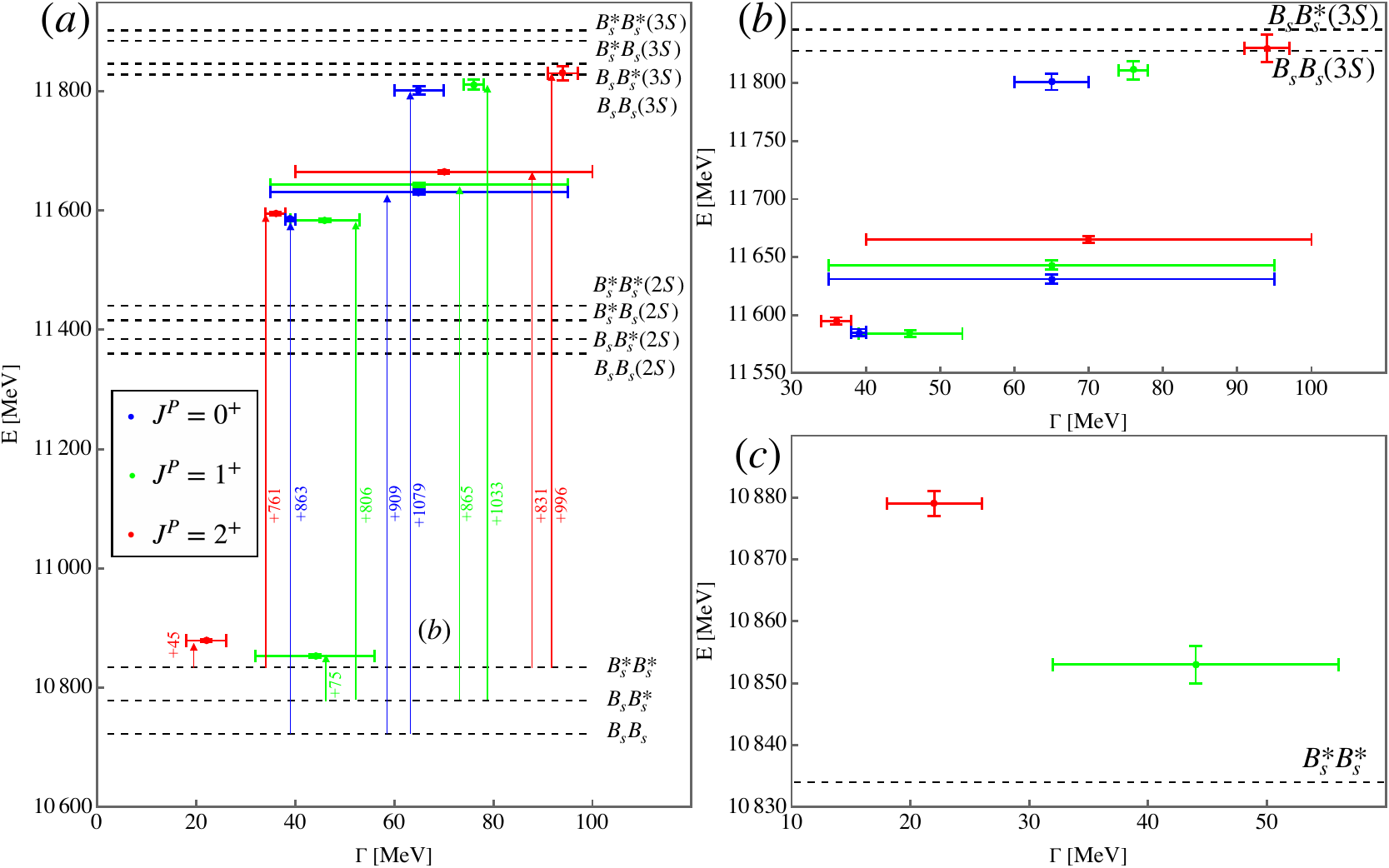}
    \caption{ The masses and widths of the $bb\bar{s}\bar{s}$ resonances obtained in the complex-scaling method and their positions relative to scattering states. For more details, see  caption of Fig. \ref{ELcccc}.}
  \label{ELbbss}
\end{figure}

\begin{figure}[h]
  \centering
  \includegraphics[width=14cm]{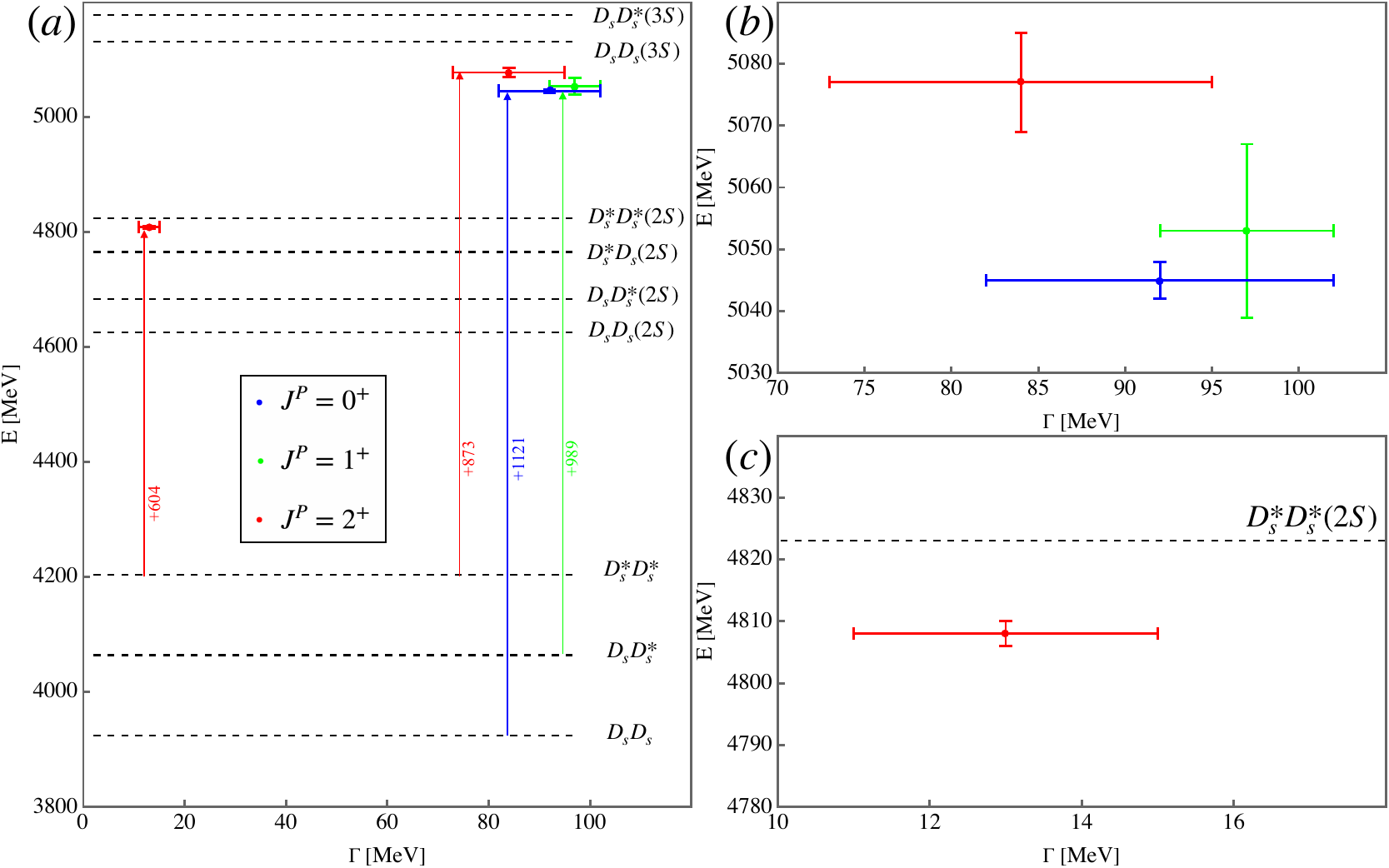}
    \caption{ The masses and widths of the $cc\bar{s}\bar{s}$ resonances obtained in the complex-scaling method and their positions relative to scattering states. For more details, see  caption of Fig. \ref{ELcccc}.}
  \label{ELccss}
\end{figure}

\begin{figure}[h]
  \centering
  \includegraphics[width=14cm]{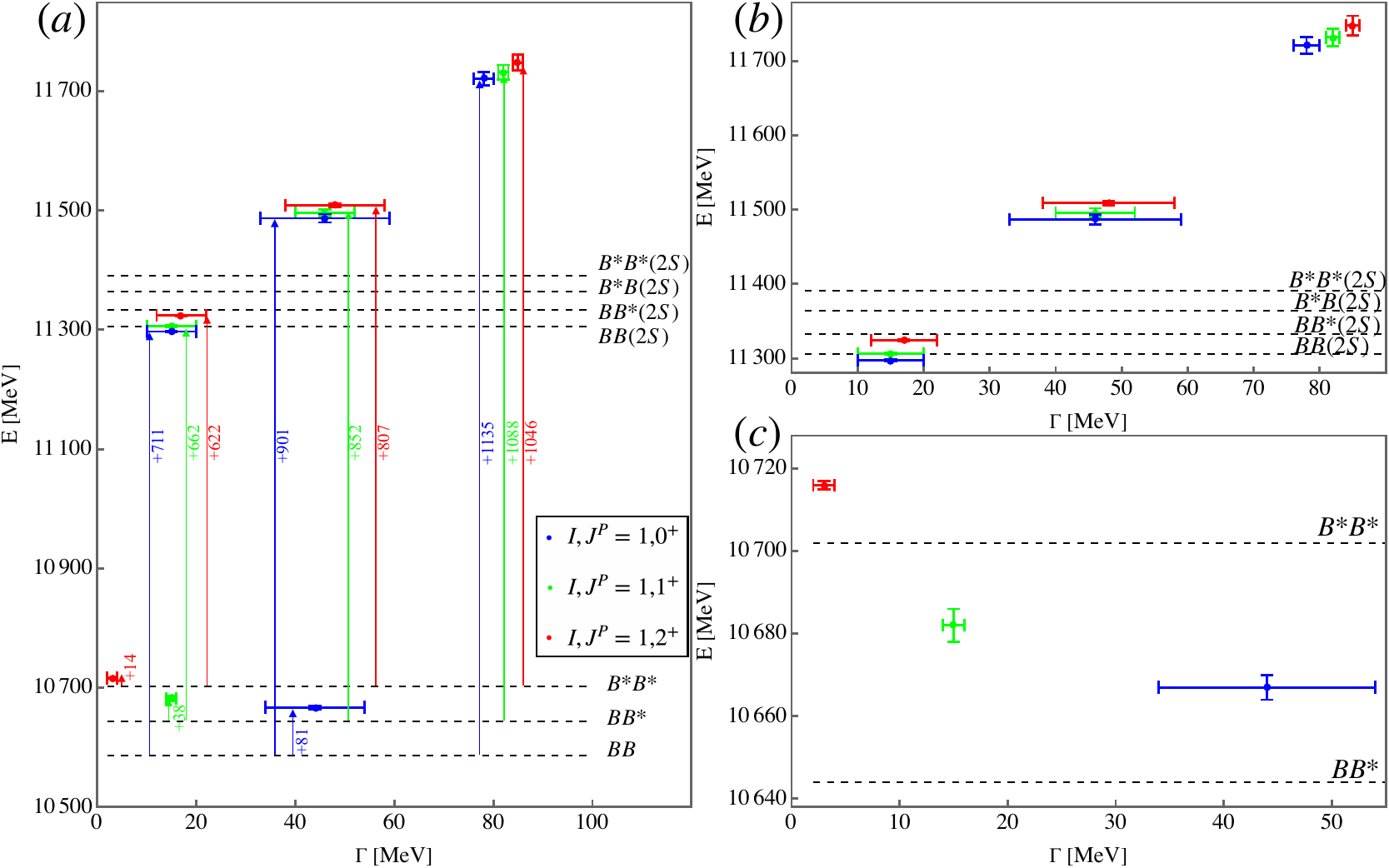}
    \caption{ The masses and widths of the $bb\bar{q}\bar{q}$ resonances with $I(J^{PC})=1(0^+)$, $1(1^+)$, $1(2^+)$ obtained in the complex-scaling method and their positions relative to scattering states. For more details, see  caption of Fig. \ref{ELcccc}.}
  \label{ELbbqq}
\end{figure}

\begin{figure}[h]
  \centering
  \includegraphics[width=14cm]{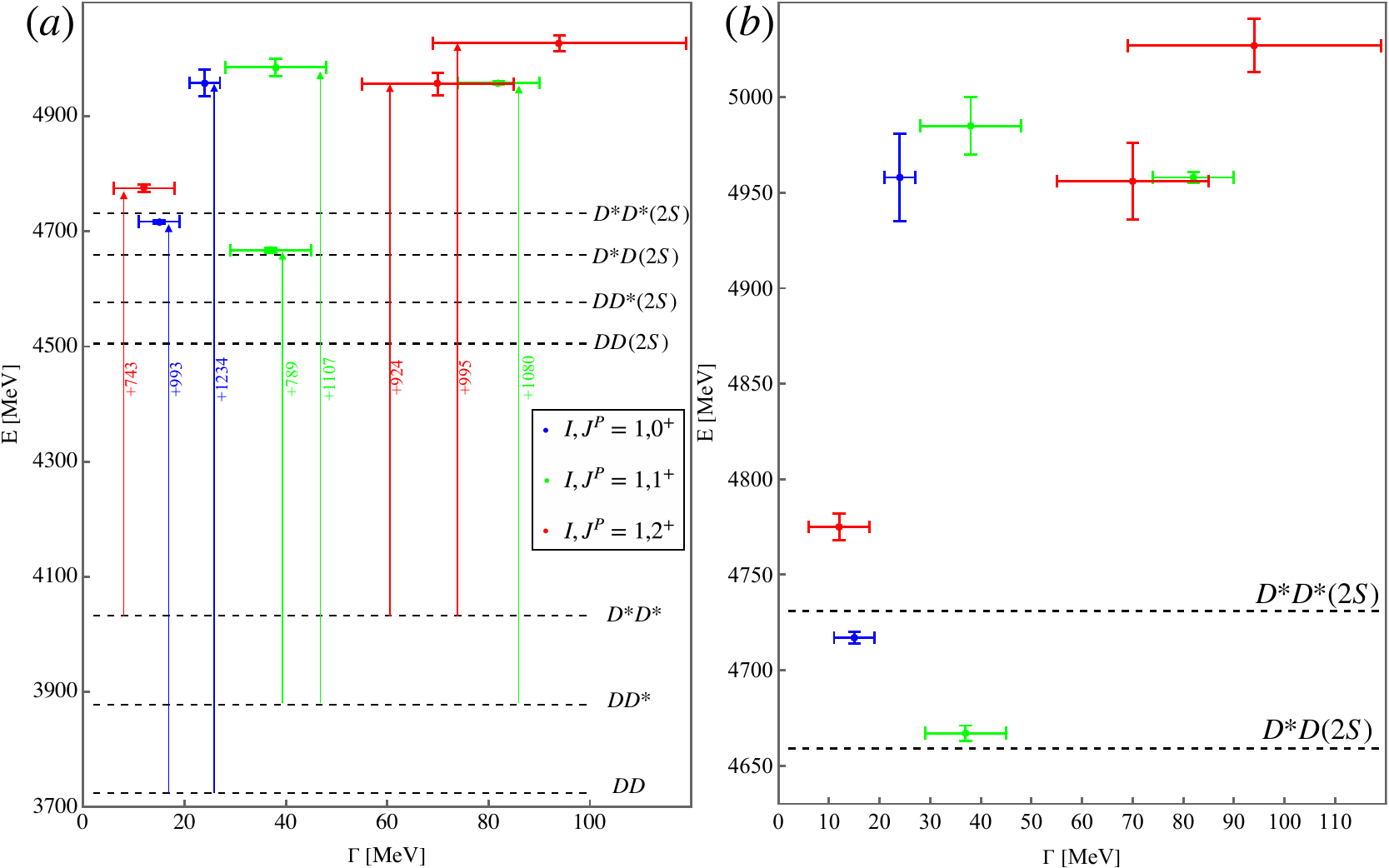}
    \caption{ The masses and widths of the $cc\bar{q}\bar{q}$ resonances with $I(J^{PC})=1(0^+)$, $1(1^+)$, $1(2^+)$   obtained in the complex-scaling method and their positions relative to low-lying scattering states. For more details, see caption of Fig. \ref{ELcccc}.}
  \label{ELccqq}
\end{figure}

\begin{figure}[h]
  \centering
  \includegraphics[width=14cm]{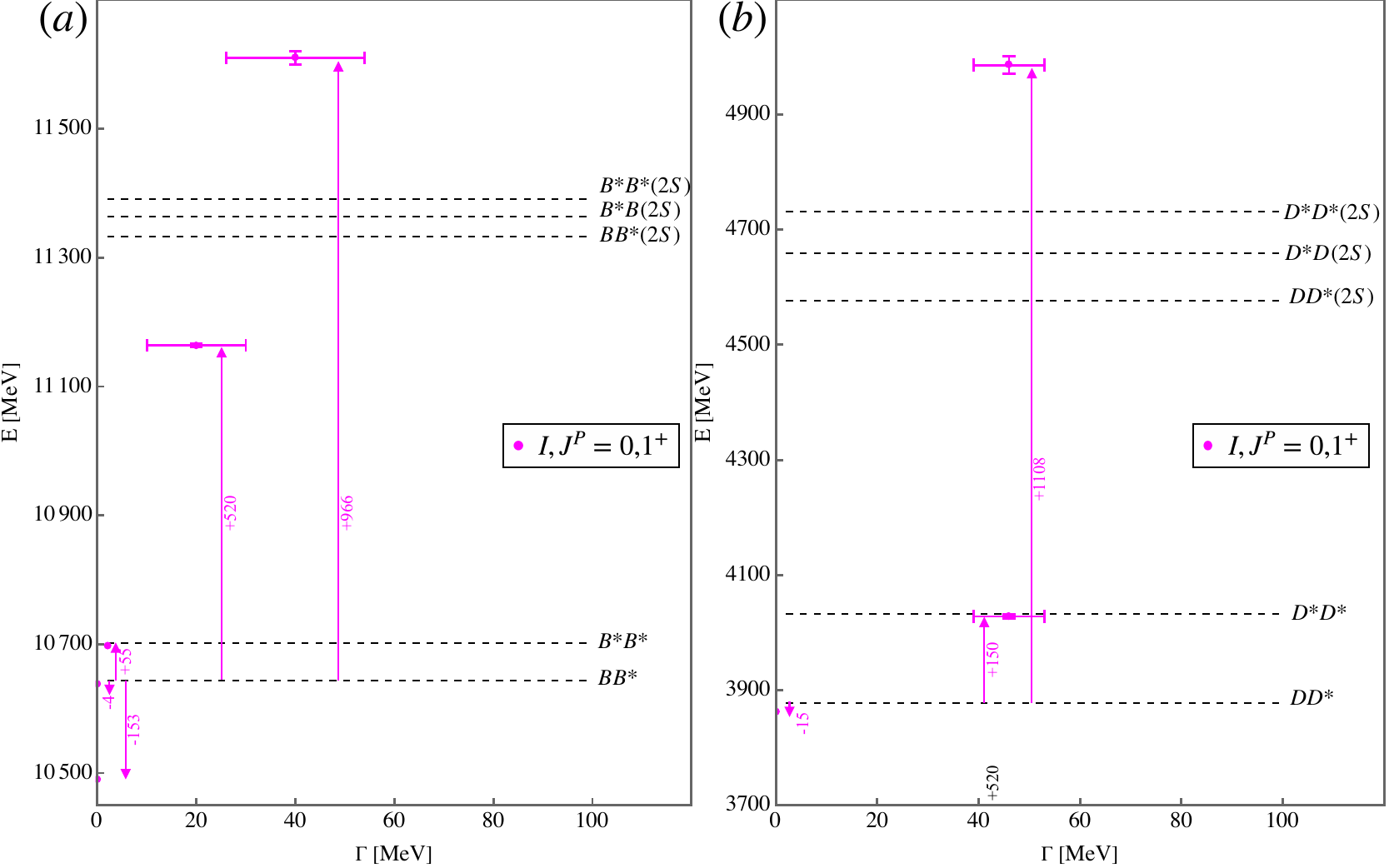}
    \caption{ The masses and widths of the (a) $bb\bar{q}\bar{q}$ and (b) $cc\bar{q}\bar{q}$ states with $I(J^{P})=0(1^{+})$ and their positions relative to the low-lying scattering threshold.}
  \label{fig:QQqq0}
\end{figure}

In this section, we present a comprehensive list of bound and resonant states of the full- and double-heavy tetraquark states using the AL1 potential in Tables \ref{tab:cccc} and \ref{tab:tetramass}. To offer a clearer perspective on the results, we illustrate the positions of these states relative to scattering states in Figs.~\ref{ELcccc}, \ref{ELbbbb}, \ref{ELbbcc}, \ref{ELbbss}, \ref{ELccss}, \ref{ELbbqq}, \ref{ELccqq} and \ref{fig:QQqq0}. Additionally, we present the complex-scaling plots that are used to identify these states in Figs.~\ref{fig:cccc}, \ref{fig:bbbb}, \ref{fig:bbcc}, \ref{fig:bbss}, \ref{fig:ccss}, \ref{fig:bbqq}, \ref{fig:ccqq} and \ref{fig:iso0csm}  in Appendix \ref{sec:ecsm}. In the CSM, we have set the scaling angle $\theta$ to range from $8^{\circ}$ to $14^{\circ}$. These complex-scaling plots clearly show that most complex eigenvalues rotate along the scattering lines, suggesting the corresponding eigenstates are predominantly ground and excited scattering states. Only the pole positions of a few states are independent of the scaling angle $\theta$, signifying these states as the bound and resonant states. 

\begin{center}
\begin{table}[h]
\caption{The masses and decay widths  $E(\Gamma)$ (in units of MeV)  of the $cc\bar{c}\bar{c}$ resonances calculated using the AL1 and BGS potentials \cite{Wang:2022yes} are presented and compared with the experimental values in LHCb \cite{LHCb:2020bwg}, CMS \cite{CMS:2023owd} and ATLAS \cite{ATLAS:2023bft} collaborations. The notations ``I" and ``II", along with ``$A$", ``$B$", ``$\alpha$" and ``$\beta$", represent  different fitting modes used in LHCb \cite{LHCb:2020bwg} and ATLAS \cite{ATLAS:2023bft}  analyses, respectively. }
\label{tab:cccc}
\begin{tabular}{c|c|c|c|c|c}
\hline \hline 
 &  & $X(6400)$ & $X(6600)$ & $X(6900)$ & $X(7200)$\tabularnewline
\hline 
\multirow{2}{*}{LHCb} & \multirow{2}{*}{MeV} & \multicolumn{2}{c|}{(6200,6800) } & $\begin{array}{c}
6905\pm 11\pm 7\\
80\pm 19\pm33
\end{array}$(I) & -\tabularnewline
 &  & \multicolumn{2}{c|}{} & $\begin{array}{c}
6886\pm11\pm11\\
168\pm 33\pm 69
\end{array}$(II) & - \tabularnewline
\hline 
CMS & MeV & - &  $\begin{array}{c}
6552\text{\ensuremath{\pm}}10\ensuremath{\pm}12\\
124\pm29\pm34
\end{array}$ & $\begin{array}{c}
6927\pm9\pm5\\
122\pm22\pm19
\end{array}$ & $\begin{array}{c}
7287\pm19\text{\ensuremath{\pm}}5\\
95\text{\ensuremath{\pm}}46\pm20
\end{array}$\tabularnewline
\hline 
\multirow{2}{*}{ATLAS} & GeV & $\begin{array}{l}
6.41\pm0.08_{-0.03}^{+0.08}\\
0.59\pm0.35_{-0.00}^{+0.12}
\end{array}(A)$ & $\begin{array}{l}
6.65\pm0.02_{-0.02}^{+0.03}\\
0.44\pm0.05_{-0.05}^{+0.06}
\end{array}(A)$ & $\begin{array}{l}
6.86\pm0.03_{-0.02}^{+0.01}\\
0.11\pm0.05_{-0.01}^{+0.02}
\end{array}(A)$ & $\begin{array}{c}
7.22\pm0.03_{-0.03}^{+0.01}\\
0.09\pm0.06_{-0.03}^{+0.06}
\end{array}(\alpha)$\tabularnewline
 &  & - & $\begin{array}{l}
6.65\pm0.02_{-0.02}^{+0.03}\\
0.44\pm0.05_{-0.05}^{+0.06}
\end{array}(B)$ & $\begin{array}{l}
6.91\pm0.01\pm0.01\\
0.15\pm0.03\pm0.01
\end{array}(B)$ & $\begin{array}{c}
6.96\pm0.05\pm0.03\\
0.51\pm0.17_{-010}^{+0.11}
\end{array}(\beta)$\tabularnewline
\hline 
\multirow{3}{*}{BGS} & $\ensuremath{0^{++}}$ & - & - & $\begin{array}{l}
7035.1(77.8)
\end{array}$ & $\begin{array}{c}
7202.2(60.6)
\end{array}$\tabularnewline
 & $1^{+-}$ & - & - & $\begin{array}{c}
7049.6(69.4)
\end{array}$ & $\begin{array}{c}
7273.5(49.8)
\end{array}$\tabularnewline
 & $2^{++}$ & - & - & $\begin{array}{c}
7068.5(83.6)
\end{array}$ & $\begin{array}{c}
7281.3(91.2)
\end{array}$\tabularnewline
\hline 
\multirow{3}{*}{AL1} & $\ensuremath{0^{++}}$ & - & - & $\begin{array}{c}
6993(84)
\end{array}$ & $\begin{array}{c}
7187(26)
\end{array}$\tabularnewline
 & $1^{+-}$ & - & - & $\begin{array}{c}
7001(70)
\end{array}$ & $\begin{array}{c}
7199(40)
\end{array}$\tabularnewline
 & $2^{++}$ & - & - & $\begin{array}{c}
7018(68)
\end{array}$ & $\begin{array}{c}
7220(40)
\end{array}$\tabularnewline
\hline \hline 
\end{tabular}
\end{table}
\end{center}

\begin{center}
\renewcommand\arraystretch{1.0} 
\begin{table}[h]
\caption{The masses and decay widths  (in units of MeV)  $E(\Gamma)$ of the resonances (bound states) calculated with AL1 potential.}
\label{tab:tetramass}
\begin{tabular}{p{1.3cm}<{\centering}|p{1.7cm}<{\centering}  |p{2.0cm}<{\centering} p{2.0cm}<{\centering}  p{2.0cm}<{\centering} p{2.0cm}<{\centering} p{2.0cm}<{\centering} p{2.0cm}<{\centering} }
\hline\hline
&$(I,)J^{P(C)}$     & \multicolumn{5}{c}{Mass(Width)}   \\
\hline
\multirow{3}{*}{$cc\bar{c}\bar{c}$}& $0^{++}$  & $6993(84)$ & $7187(26)$ &   &  \\
& $1^{+-}$  & $7001(70)$ & $7199(40)$ &  &    \\
& $2^{++}$  & $7018(68)$ & $7220(40)$ &  &    \\
\hline
\multirow{3}{*}{$bb\bar{b}\bar{b}$}& $0^{++}$  & $19790(58)$ & $19960(24)$ &  &    \\
& $1^{+-}$  & $19794(58)$ & $19960(28)$ &  &    \\
& $2^{++}$  & $19800(60)$ & $19963(31)$ &  &    \\
\hline
\multirow{3}{*}{$bb\bar{c}\bar{c}$}& $0^{+}$  & $13449(74)$ & $13585(26)$ & $13680(26)$ &  \\
& $1^{+}$  & $13451(68)$ & $13568(18)$ & $13684(30)$ &  &  \\
& $2^{+}$  & $13462(72)$ & $13560(12)$ & $13652(24)$ & $13685(50)$ &  \\
\hline
\multirow{3}{*}{$bb\bar{s}\bar{s}$}
& $0^{+}$  & $11585(39)$ & $11631(65)$ & $11801(65)$ &   \\
& $1^{+}$  &$10853(44)$& $11584(46)$ & $11643(65)$ & $11811(76)$ &    \\
& $2^{+}$  &$10879(22)$& $11595(36)$ & $11665(70)$ & $11830(94)$ &    \\
\hline
\multirow{3}{*}{$cc\bar{s}\bar{s}$}
& $0^{+}$  & $5045(92)$ &  &  &    \\
& $1^{+}$  & $5053(97)$ &  &  &    \\
& $2^{+}$  & $4808(13)$ & $5077(84)$ &  &  \\
\hline
\multirow{4}{*}{$bb\bar{q}\bar{q}$}
& $1,0^{+}$  & $10667(44)$ & $11297(15)$ & $11487(46)$ & $11721(78)$ &  \\
& $1,1^{+}$  & $10682(15)$ & $11306(15)$ & $11496(46)$ & $11732(82)$ &   \\
& $1,2^{+}$  & $10716(3)$ & $11324(17)$ & $11509(48)$ & $11748(85)$ &   \\
& $0,1^{+}$  & $10491$ & $10640$ & $10699(2)$ & $11164(20)$ & $11610(40)$   \\
\hline
\multirow{4}{*}{$cc\bar{q}\bar{q}$}
& $1,0^{+}$  & $4717(15)$ & $4958(24)$ &  &  &   \\
& $1,1^{+}$  & $4667(37)$ & $4958(82)$ & $4985(38)$ &  &   \\
& $1,2^{+}$  & $4775(12)$ & $4956(70)$ & $5027(94)$ &  &   \\
& $0,1^{+}$  & $3863$ & $4028(46)$ & $4986(46)$ &  &  &  \\
\hline\hline
\end{tabular}
\end{table}
\end{center}

\subsection{The $cc\bar{c}\bar{c}$ tetraquark} \label{subsec:cccc}

We initially employ the AL1 potential to calculate the $cc\bar{c}\bar{c}$ spectra, with the results summarized in Table~\ref{tab:cccc} and Fig. \ref{ELcccc}. These results are then compared with those obtained using the BGS potential in our previous work \cite{Wang:2022yes}.

No bound states are observed in $J^{PC}=0^{++}$, $1^{+-}$ and $2^{++}$ sectors with  AL1 potential, which are consistent with the results obtained from the BGS potential \cite{Wang:2022yes}. Additionally, we identify totally $6$ $cc\bar{c}\bar{c}$ resonances with $2$ resonances in each $J^{PC}$ quantum number category.  The $0^{++}$ states have a lower mass than  the $1^{+-}$ and $2^{++}$ states. Owing to the heavy quark symmetry, the hyperfine interactions are suppressed, leading to similar resonance spectra across $J^{PC}=1^{+-}$ and $2^{++}$ sectors. The lower resonances are located around $7000$ MeV, while the higher ones are found at approximately $7200$ MeV.

Next, we will delve into the specifics of each resonant state.
For $J^{PC}=0^{++}$: The lower resonance is located at $E=6993$ MeV with a width of $\Gamma=84$ MeV, and potentially decay into the $S$-wave  di-$\eta_c$, di-$J/\psi$, $\eta_c \eta_c(2S)$ and $J/\psi \psi(2S)$ channels. The higher resonance,  located at $E=7187$ MeV with a width of $\Gamma=26$ MeV, may decay into the di-$\eta_c$, di-$J/\psi$, $\eta_c \eta_c(2S)$, $J/\psi \psi(2S)$, $\eta_c \eta_c(3S)$ and $J/\psi \psi(3S)$ channels.  

For $J^{PC}=1^{+-}$: Two resonances which are identified at $E=7001$ MeV with a width of $\Gamma=70$ MeV and $E=7199$ MeV with a width of $\Gamma=40$ MeV. The lower one could decay into the $\eta_c J/\psi$, $\eta_c \psi(2S)$, and $J/\psi \eta_c(2S)$ channels while the higher one into the $\eta_c J/\psi$, $\eta_c \psi(2S)$, $J/\psi \eta_c(2S)$, $\eta_c \psi(3S)$, and $J/\psi \eta_c(3S)$ channels.

For $J^{PC}=2^{++}$: we obtained two resonances with $E=7018$ MeV ($\Gamma=68$ MeV) and $E=7220$ MeV ($\Gamma=40$ MeV). The lower one may decay into the di-$J/\psi$, and $J/\psi \psi(2S)$ channels, while the higher one  into the di-$J/\psi$, $J/\psi \psi(2S)$ and $J/\psi \psi(3S)$ channels. They are also the candidates for the $X(6900)$ and $X(7200)$ in experiments.

For $cc\bar{c}\bar{c}$ system, results obtained with the AL1 and BGS potentials are generally consistent as shown in Table \ref{tab:cccc}, though the BGS predictions are about tens of MeV (up to $60$ MeV) higher than those of AL1. This discrepancy is consistent with expectations based on the charmonium spectrum, as shown in Table \ref{tab:hm}. Specifically, the excited $c\bar c$ states calculated in BGS are up to $30$ MeV larger than those calculated with AL1. Therefore, a similar trend of smaller masses for the excited tetraquark states in the AL1 model is expected.

All the $cc\bar c\bar c$ states, including $X(6600)$, $X(6900)$ and $X(7200)$ have been observed in the di-$J/\psi$ or $J/\psi \psi(2S)$ channel in experiments \cite{LHCb:2020bwg, CMS:2023owd,ATLAS:2023bft}.  The $J^{PC}$ quantum numbers have not been assigned so far for the available experimental data.  When these states are considered as the S-wave tetraquark states, their $J^{PC}$ may be either $=0^{++}$ or $2^{++}$. Compared with experimental data, the AL1 potential predicts a mass for $X(6900)$  that is lower than that of the BGS potential, but still approximately $100$ MeV heavier than the experimental value. However, the decay width obtained from BGS and AL1 are similar and consistent with experimental values, considering the experimental uncertainties. For $X(7200)$ resonance, the AL1 predicts a smaller mass than BGS. The masses and decay widths from both models are consistent with the experimental results for $X(7200)$ taking into account the uncertainties. However, AL1 predicts a smaller decay width compared to BGS potential. It should be noted that the ATLAS has also proposed an alternative interpretation \cite{ATLAS:2023bft}. Their analysis suggests that the highest pole may correspond to a resonance located at $6.96\pm0.05\pm0.03$ GeV by using a single resonance fit to their $J/\psi\psi(2S)$ data. Therefore, it is not currently possible to determine the $J^{PC}$ of $X(6900)$ and $X(7200)$ based solely on the masses and widths. We expect more detailed and precise experimental data in the future to further investigate these fully charmed tetraquark states.

Both AL1 and BGS potentials fail to predict any resonant state in the energy region of $6.2 - 6.8$ GeV,  which is where the experimentally suggested $X(6400)$ and $X(6600)$ resonances might be located.  It is important to note that both models employ a confinement potential characterized as the summation of quark-(anti)quark linear confinement. The alteration of the confinement mechanism has led to the appearance of the bound states and a lower $X(6600)$, as demonstrated in our previous work \cite{Wang:2023jqs}. These findings can serve as indicators for analyzing the confinement mechanism in the multiquark states.

Finally, it is important to emphasize the potential existence of even higher resonances. However, in CSM, these highly excited states exhibit strongly oscillating asymptotic wave functions over long distances. This makes them challenging to describe accurately and results in bad convergence or even the disappearance of the resonance poles in our analysis. Additionally, for these higher states, the other excited scattering states with higher orbital excitations may contribute importantly and cannot be neglected. Therefore, we primarily focus on the low resonant states both for the $cc\bar c\bar c$ states and in subsequent analyses.

\subsection{$bb\bar{b}\bar{b}$ and $bb\bar{c}\bar{c}$}

Experiments have not yet observed any  $bb\bar{b}\bar{b}$ states in the $\Upsilon(1S)\mu^+\mu^-$ channels and the experimental data for  $bb\bar{c}\bar{c}$ states is absent. For the AL1 potential, we also do not observe any bound states. These findings are consistent with Lattice QCD calculations, which also have not found the bound $bb\bar{b}\bar{b}$ \cite{Hughes:2017xie} and $bb\bar{c}\bar{c}$ states \cite{Junnarkar:2018twb}. On the other hand, we identified $16$ resonant states for the $bb\bar{b}\bar{b}$ and $bb\bar{c}\bar{c}$ states, and the results are summarized in Table \ref{tab:tetramass} and Figs. \ref{ELbbbb} and \ref{ELbbcc}.

For the $bb\bar{b}\bar{b}$ system, we have obtained six resonances: two each for $J^{PC}=0^{++}$, $1^{+-}$, and $2^{++}$. These resonances are analogous to  the $cc\bar{c}\bar{c}$ resonances, with  $c$ and $\bar{c}$ quarks are substituted by $b$ and $\bar{b}$ quarks.  Notably, the heavy quark spin symmetry is more pronounced in $bb\bar b\bar b$ states compared to the $cc\bar c\bar c$ states. The two resonances within each of the three $J^{PC}$ quantum numbers exhibit similar widths and positions. The lower resonances are situated around $19790$ MeV and have the decay widths in the order $\mathcal O (60)$ MeV. The higher resonances are located around $19960$ MeV with narrower widths of approximately $\mathcal O(30)$ MeV. We expect that more data in the lower scattering channels, such as di-$\Upsilon(1S)$, $\Upsilon(1S)\Upsilon(2S) $, and possibly $\Upsilon(1S)\Upsilon(3S)$ channels in future to help the search for these resonances.

For the $bb\bar{c}\bar{c}$ ($\bar b \bar b {c}{c}$) system, we totally identified $10$ resonances: three for $J^{P}=0^{+}$, three for $J^{P}=1^{+}$, and four for $J^{P}=2^{+}$. The $bb\bar{c}\bar{c}$ system is similar to the $bb\bar{b}\bar{b}$ or $cc\bar{c}\bar{c}$ systems when substituting $c$ ($\bar{b}$) quarks with $b$ ($\bar{c}$) quarks. However,  the lowest two resonances in the $bb\bar{c}\bar{c}$ system, are likely closer to the lowest scattering states compared to their counterparts in the $bb\bar{b}\bar{b}$ and $cc\bar{c}\bar{c}$ systems. The primary reason for this is the mass difference between the $b$ and $c$ quarks, as discussed in Sec. \ref{quark mass ratio}.

In the heavy quark limit, the spectra of $1^+$ and $2^+$ states are expected to be similar. However, the channel couplings to the scattering states make them different.
The $2^+$ states couples only to the $B_c^*B_c^*$ scattering state in the $S$ wave. This results in more straightforward patterns of the eigenvalues in the complex energy plane. The $1^{+}$ sector features a more complex scenario due to a larger number of scattering states within the concerned energy range of $12600$ MeV to $13600$ MeV. This complexity is caused by the small splittings between $B_c$ and $B_c^*$.

\subsection{Double-heavy tetraquark states $QQ\bar s\bar s$ ($QQ \bar s \bar s$) and $QQ\bar q\bar q$ ($\bar Q\bar Q qq$) with isospin $I=1$}

The $QQ\bar{s}\bar{s}$  and $QQ\bar{q}\bar{q}$ states with isospin $I=1$ have the identical color-spin-flavor configurations with the fully-heavy tetraquark states $bb\bar{c}\bar{c}$.  The possible $J^{P}$ quantum numbers of the S-wave tetraquark states are $0^+$,  $1^{+}$ and $2^{+}$. Notably, the double-heavy tetraquark states have larger mass ratios of the quark $Q$ and the antiquark $\bar q$ or $\bar s$, $m_Q/m_{\bar q/\bar s}$, compared to the  $bb\bar{b}\bar{b}$, $cc\bar{c}\bar{c}$ and $bb\bar{c}\bar{c}$ tetraquarks. 
It was suggested that the larger mass ratio can facilitate the formation of stable tetraquark states or low-energy tetraquark resonances~\cite{Karliner:2017qjm,Eichten:2017ffp}.

We present the spectra of  $bb\bar{s}\bar{s}$ and $cc\bar{s}\bar{s}$ calculated using the AL1 potential in Table~\ref{tab:tetramass} and the pole positions relative to the scattering states are displayed in Figs. \ref{ELbbss}, \ref{ELccss}. No bound states are observed. In the case of $bb\bar{s}\bar{s}$, we totally identified $9$ resonances below the $B^{*}_sB_s(3S)$ threshold: three each for $J^{P}=0^{+}$, $J^{P}=1^{+}$ and $J^{P}=2^{+}$. Similarly, for $cc\bar{s}\bar{s}$, we identified the  $4$ resonances below the $D^{(*)}_sD^*_s(3S)$ threshold: one for $J^{P}=0^{+}$, one for $J^{P}=1^{+}$ and two for $J^{P}=2^{+}$.

The results of isospin $I=1$ $bb\bar{q}\bar{q}$ and $cc\bar{q}\bar{q}$ spectra calculated using the AL1 potential are displayed in Figs.~\ref{ELbbqq}, \ref{ELccqq}. We do not find bound states  for isovector $bb\bar q\bar q$ systems but  $12$ resonances: four each for the states with $I(J^{P})=1(0^{+})$, $I(J^{P})=1(1^{+})$ and $I(J^{P})=1(2^{+})$.

For $I(J^{P})=1(0^{+})$, the lowest resonance is at $E(\Gamma)=10667(44)$ MeV. It is about $81$ MeV above the $BB$ threshold and roughly $35$ MeV below the $B^*B^*$ threshold. The most important decay channel is into $BB$. The second resonance, $E(\Gamma)=11297(15)$ MeV is  only around $10$ MeV below the $BB(2S)$ threshold and  may decay into both $BB$ and $B^*B^*$ channels. The two higher ones may decay into the S-wave $BB$, $B^*B^*$,  $BB(2S)$, and $B^*(2S)B^*$ . 

For $I(J^{P})=1(1^{+})$, two lower states  $E(\Gamma)=10682(15)$ MeV and $E(\Gamma)=11306(15)$ MeV may decay into $BB^*$ channels. The higher two resonances $E(\Gamma)=11496(46)$ MeV and $E(\Gamma)=11732(82)$ MeV may decay into $BB^*$, $BB^*(2S)$, $B^*B(2S)$ and $B^*B^*(2S)$ channels. 

For $I(J^{P})=1(2^{+})$, the two lower resonances, $E(\Gamma)=10716(3)$ MeV and $E(\Gamma)=11324(17)$ MeV may decay into $B^*B^*$ threshold. $E(\Gamma)=11509(48)$ MeV and $E(\Gamma)=11748(85)$ MeV may decay into $B^*B^*$ and $B^*B^*(2S)$. To be noticed, 
the lowest resonance $E(\Gamma)=10716(3)$ MeV is located at $\sim14$ MeV above $B^*B^*$ threshold and with a decay width of only $3$ MeV. 

In a previous study by Meng et al. using a different quark model \cite{Meng:2020knc}, a bound state of $I(J^{P})=1(2^{+})$ $bb\bar{q}\bar{q}$ was identified, located at $4$ MeV below the $B^*B^*$ threshold. This discrepancy of about $18$ MeV between the two models highlights the model dependency of the states near the threshold. In addition, the impact of the long-range potential, such as light meson exchange potential may be crucial and cannot be neglected near the threshold.
Using the same quark model potential, Meng et al. calculated the resonant states using the real-scaling method in another study  \cite{Meng:2021yjr}, observing a resonance at $E(\Gamma) = 10641 (15)$ MeV for $I(J^{P})=1(1^{+})$  $bb\bar q\bar q$. The result agrees with our observation of a resonance at  $E(\Gamma)=10682(15)$ MeV. Although the mass is $41$ MeV higher in our study, the decay width is similar, and the relative distance from the thresholds is close, further illustrating the potential variations and model-dependent nature of the calculations. 

For $I=1$ $cc\bar{q}\bar{q}$ system, while no bound states were observed, we identified a total of $8$ resonances: two for $I(J^{P})=1(0^{+})$, three resonances each for $I(J^{P})=1(1^{+})$ and  $I(J^{P})=1(2^{+})$. 

Comparing $bb\bar s\bar s$ and $cc\bar s\bar s$ as well as the $I=1$ $bb\bar q\bar q$ and $cc\bar q\bar q$ states, respectively,  we note that the lowest resonant states are present close to the corresponding lowest scattering states in the bottom sector. However, the analogous states disappear in the charmed sector. 
As Ref. \cite{Meng:2021yjr} suggests, this could be because of the increased short-range attractive potentials in the $QQ'\bar{q}\bar{q}$ system with a larger reduced mass of $QQ'$. Compared with the $bb$ system, the weaker attractive potential in $cc $ leads to a less compact $c c \bar{q} \bar{q}$ tetraquark,  suggesting stronger coupling to the $D^{(*)}D^{(*)}$ channels due to the larger spatial overlap. This makes the analogous lowest states in the charmed sector more likely to be predominantly scattering states than resonances.

\subsection{Double-heavy tetraquark states $QQ\bar q\bar q$ ($\bar Q\bar Q qq$) with isospin $I=0$}

This section explores the intriguing heavy tetraquark states $QQ\bar q\bar q$ ($\bar Q\bar Q qq$) with isospin $I=0$. Constrained by the Fermi-Dirac statistics, the S-wave tetraquark states  $QQ\bar q\bar q$ with $I=0$ are limited to $I(J^P)=0(1^+)$. Experimentally, the $T_{cc}$ is the only observed double-charmed state.  While $T_{bb}$ state has not been observed, all the lattice QCD calculations suggest its existence, albeit with differing binding energies ranging from $-59 \pm 38$ MeV  \cite{Bicudo:2016ooe}  to  $ -189 \pm 13 $ MeV  \cite{Francis:2016hui} as summarized in review \cite{Bicudo:2022cqi}. The study of the  $QQ\bar q\bar q$ ($\bar Q\bar Q qq$) state will help to investigate the nature of the observed $T_{cc}$ state and help experimental exploration of the $T_{bb}$ state. The obtained results are presented in Table \ref{tab:tetramass} and Fig.~\ref{fig:QQqq0}.

For the $I(J^P)=0(1^+)$ $bb\bar{q}\bar{q}$ state, we have identified two bound states and three resonances. The first bound state at $10491$ MeV has a deep binding energy of $153$ MeV to the $BB^*$ threshold and is below the $BB$ threshold, suggesting the possible decay only through weak channels. The second bound state appears at $10640$ MeV, slightly under the $BB^*$ threshold by $4$ MeV but above the $BB$ threshold. The $1^+$ state cannot decay into $BB$ via the strong interaction. It decays only radiatively to $BB\gamma$. These findings are consistent with the earlier results using a different potential \cite{Meng:2020knc}. The lattice QCD calculations supported the existence of the lower state with a large binding energy,  but not necessarily the second bound state near the threshold. Caution is warranted for the second bound state. Phenomenologically, the long-range interactions for the near-threshold states may change the situation.  Thus the second bound state might shift or even disappear due to the long-range interactions, such as the light-mesons exchange. In lattice QCD, the $T_{bb}$ state is computed using various approaches to treat the heavy $b $ quark, including static  \cite{Bicudo:2016ooe,Bicudo:2012qt,Bicudo:2015vta,Bicudo:2021qxj}, heavy quark  \cite{Francis:2016hui,Francis:2018jyb,Hudspith:2020tdf,Junnarkar:2018twb,Leskovec:2019ioa,Colquhoun:2022dte} or scattering lattice QCD \cite{Aoki:2022xxq,Wagner:2022bff,Pflaumer:2022lgp}. The uncertainties of the results are in the tens of MeV range. Therefore, the second bound state near the threshold should be examined with additional scrutiny.

The lowest resonance is at $10699$ MeV with a narrow width of $2$ MeV, located $3$ MeV below the $B^*B^*$ threshold. It is mostly likely to be observed in the $BB^*$ channel.  Both the mass and width are consistent with the previous calculation using the real-scaling method \cite{Meng:2021yjr}, where the resonant state is interpreted as a mixture of the $bb(2S)\bar{q}\bar{q}$ tetraquark and $B^*B^*$ meson-meson state.
In addition, we have identified two additional resonances, which were not covered in the energy range studied in Ref. \cite{Meng:2021yjr}. The first, $E(\Gamma)=11164(20)$ MeV is  $ 520$ MeV above the lowest threshold $BB^*$ threshold and about $170$ MeV below the $BB^*(2S)$ threshold. the second one with $E(\Gamma)=11610(40)$ MeV is located above the $B^*B^*(2S)$ threshold. Their convergence in CSM is less robust compared to the two bound states and the lowest resonance, suggesting more complex coupled channel effects that challenge their description using the CSM.

For $I(J^P)=0(1^+)$ $cc\bar{q}\bar{q}$, we have obtained one bound state and two resonances. The bound state at $3863$ MeV is located  $15$ MeV below the $DD^*$ threshold. This finding is consistent with the $-23$ MeV binding energy in Ref. \cite{Meng:2021yjr}. This state is significantly deeper than the experimental $T_{cc}$ state \cite{LHCb:2021vvq,LHCb:2021auc} with a unitarized binding energy at $-361 \pm 40$ keV and it is even below the  $DD\pi$ threshold. It cannot be observed in the $DD\pi$  invariant mass spectrum, however, may be observable in the  $DD\gamma$ channels.  Therefore, this finding does not support the experimental $T_{cc}$  as a tetraquark state. 
The lower resonance, $E(\Gamma)=4028(46)$ MeV, is $4$ MeV below the $D^*D^*$ threshold which may  be analogous to the low-energy resonance $E(\Gamma)=10669(2) $ MeV in $bb\bar{q}\bar{q}$ when $b$ quarks are replaced by $c$ quarks. Similar to the $bb\bar{q}\bar{q}$ state, we conjecture that the $cc\bar q\bar q$ state with $E(\Gamma)=4028(46)$ MeV is likely a mixture of the  $cc(2S)\bar{q}\bar{q}$ tetraquark and $D^*D^*$ meson-meson state. The smaller color-electric attraction in $cc$ diquark leads to a less compact $cc(2S)\bar{q}\bar{q}$ tetraquark and a larger width. The higher resonance, $E(\Gamma)=4986(46)$ MeV located above the $D^*D^*(2S)$ threshold. Both the two resonant states can be explored in the $D^*D$ and $DD\pi$ channels.

\begin{center}
\renewcommand\arraystretch{1.3} 
\begin{table}[h]
\caption{The percentages of the compact diquark-antidiquark $(P_{\text{di}})$ and meson-meson ($P_{MM^*}$/$P_{M^*M^*}$) configurations in the calculated (Cal.)  bound states alongside the comparisons with the Lattice QCD (LQCD) results \cite{Bicudo:2021qxj}. }
\label{tab:percentage}
\begin{tabular}{p{1.5cm}<{\centering}|p{2.5cm}<{\centering}  p{2.0cm}<{\centering}| p{1.5cm}<{\centering} p{1.5cm}<{\centering} p{1.5cm}<{\centering}  }
\hline\hline
  & state & $\Delta E $ [MeV] & $P_{\text{di}}$ & $P_{MM^*}$ & $P_{M^*M^*}$\\
\hline
 LQCD &$bb\bar q \bar q$ $0(1^+)$   &-$38\pm 18$ & $40\%$ &\multicolumn{2}{c}{$60\%$}  \\
 \hline
 \multirow{3}{*}{Cal.} & $bb\bar q \bar q$ $0(1^+)$ &  -153  & $63.9\%$ & $24.1\%$ & $12.0\%$ \\
  & $bb\bar q \bar q$ $0(1^+)$ &  -4  & $1.9\%$ & $94.2\%$ & $3.9\%$ \\
  & $cc\bar q \bar q$ $0(1^+)$ &  -15  & $27.3\%$ & $72.4\%$ & $0.3\%$ \\
\hline\hline
\end{tabular}
\end{table}
\end{center}

Finally, we present the percentages of the diquark-antidiquark and meson-meson components in Table \ref{tab:percentage}. It is observed that as binding energy increases, the diquark-antidiquark component becomes more important. In the deeply bound $bb\bar q\bar q$ state, the diquark-antidiquark component dominates, contributing to $63.9\%$. Conversely, for the shallower  $bb\bar q\bar q$ and $cc\bar q\bar q$ states, the meson-meson components are more important, contributing to $98.1\%$ and $72.7\%$.

\subsection{Heavy quark} \label{quark mass ratio}
\begin{figure}[h]
  \centering
  \includegraphics[width=12cm]{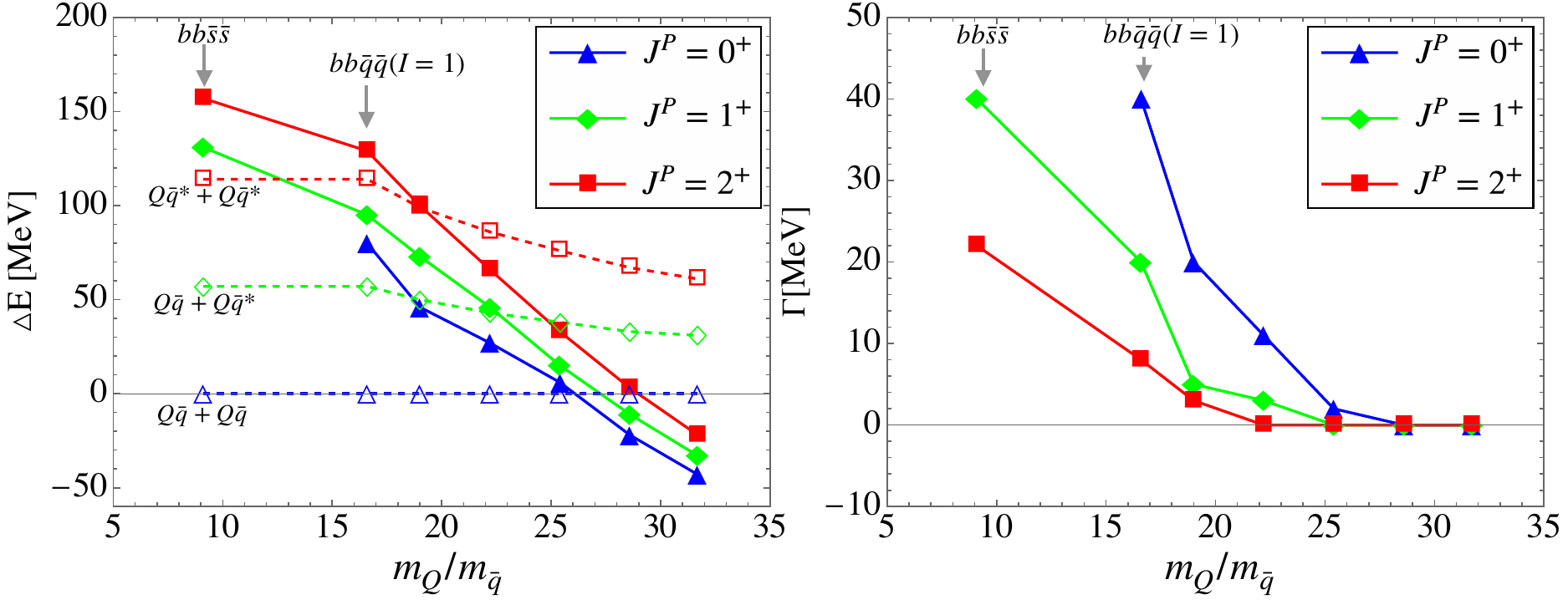}
    \caption{The relationship between the $m_Q/m_{\bar{q}}$ ratio and the pole positions of lowest-lying $QQ\bar q\bar q$ states with isospin $I=1$ is shown: (left panel) shows mass differences ($\Delta E$) from threshold $Q\bar q + Q\bar q$, and (right panel) shows decay widths ($\Gamma$).  The colored dashed lines represent the thresholds for two heavy mesons scattering states $Q{\bar q}^{(*)} + Q{\bar q}^{(*)}$  where the notation with/without the superscript $``*"$ indicates that the spin of $Q\bar q$ as $1$/$0$.}
  \label{massratio}
\end{figure}

In Fig.~\ref{massratio}, we illustrate how the pole positions of the lowest-lying states including the mass differences from the thresholds and the decay width depend on the $m_Q/m_{\bar{Q}(\bar{q})}$ ratio. We start with the  physical  $bb\bar s\bar s$ and $I=1$ $bb\bar q\bar q$  states  with the  ratios  $m_b/m_{\bar s}= 9.1$ and $m_b/m_{\bar q}= 16.6$, respectively. Then we keep the $m_{\bar q}$ invariant and increasing  $m_Q$.

One observes that as  $m_Q/m_{\bar{Q}(\bar{q})}$ ratio increases, both the masses and decay widths of the resonances systematically decrease, transitioning towards bound states. The binding energy rises  with increasing $m_Q/m_{\bar{Q}(\bar{q})}$ ratio. At a ratio of $\sim 20$  ($QQ\bar q\bar q$ with $m_Q\sim 6.3$ GeV), $J^P=2^+$ state becomes bound. The $J^P=1^+$ and $0^+$ states become bound at at ratios of $\sim 23$ ( $m_Q\sim 7.2$ GeV) and $\sim26$ ($QQ\bar q\bar q$ with $m_Q\sim 8.3$ GeV), respectively. 

This tendency was pointed out in early studies \cite{Ader:1981db,Heller:1985cb,Semay:1994ht,Vijande:2007ix,Richard:2022fdc}. They found that if $m_Q/m_{\bar{Q}(\bar{q})}$ ratio is significant, the color-electric effects $ V^{ce}$ can provide attraction for the $\bar 3_c$ diquark, lead to a $QQ\bar q\bar q$  bound state. Later findings suggest that a larger mass ratio is needed if only color-electric forces are considered.  However, adding color-magnetic forces  $ V^{cm} $, which provide attraction for the color $\bar 3_c$ $qq$ diquark with spin $S=0$, will combine with the color-magnetic forces to form a bound state with a real mass ratio \cite{Janc:2004qn,Lee:2009rt}.

\begin{center}
\renewcommand\arraystretch{1.2} 
\begin{table}[h]
\caption{The contributions of the color-electric interaction $\langle V_{bb}^{ce} \rangle$ and color-magnetic interaction $\langle V_{\bar{q}\bar{q}}^{cm} \rangle$ for $I(J^P)=0(1^+)$ and $1(1^+)$ $bb\bar q \bar q$ states. $\Delta E$ is the mass measured from the $BB^*$ threshold.}
\label{tab:contribution}
\begin{tabular}{p{1.5cm}<{\centering}|p{1.5cm}<{\centering}  p{1.5cm}<{\centering} p{1.5cm}<{\centering} p{1.5cm}<{\centering}  }
\hline\hline
$I(J^P)$  & $E$ & $\Delta E$ & $\langle V_{bb}^{ce} \rangle$ & $\langle V_{\bar{q}\bar{q}}^{cm} \rangle$ \\
\hline
 $0(1^+)$ & 10491 & -153 & -231.2 & -214.3 \\
 $1(1^+)$ & 10682 &  +38 & -102.7 & +19.2  \\
\hline\hline
\end{tabular}
\end{table}
\end{center}

\begin{figure}[h]
  \centering
  \includegraphics[width=8cm]{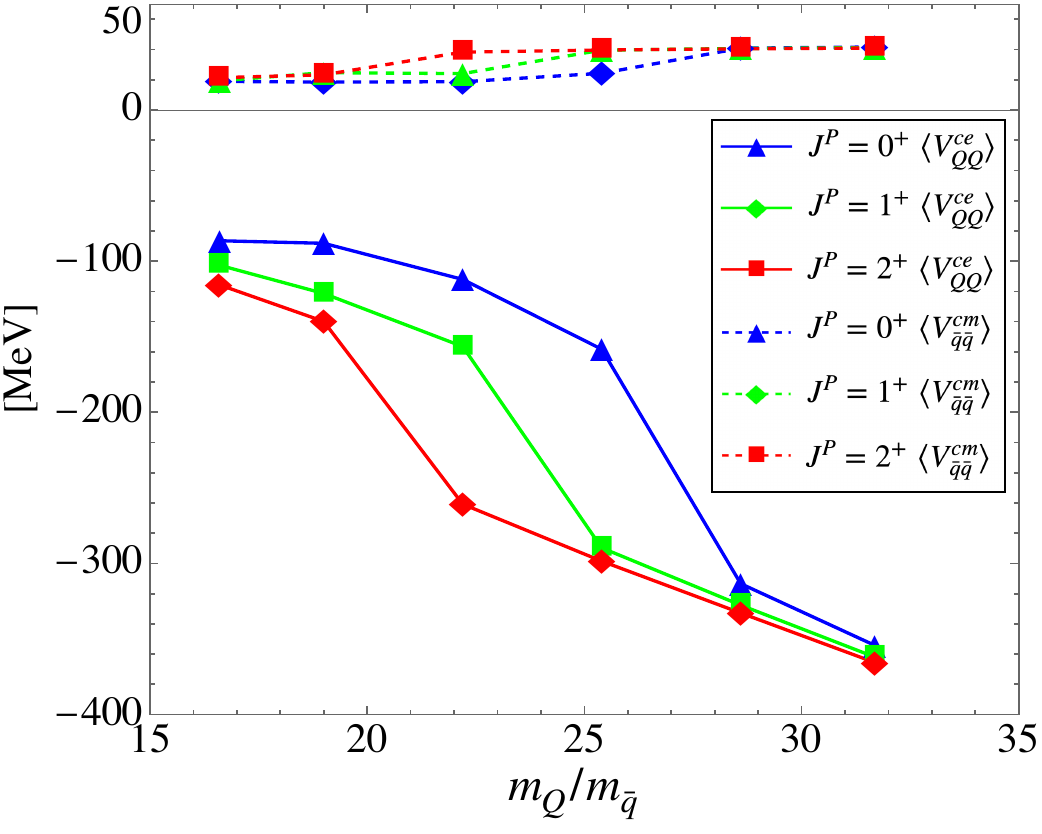}
    \caption{The $m_Q/m_{\bar{Q}(\bar{q})}$ ratio dependence of $QQ$ color-electric interaction $\langle V_{QQ}^{ce} \rangle$ and $\bar q\bar q$ and color-magnetic interaction $\langle V_{\bar{q}\bar{q}}^{cm} \rangle$ in the $QQ \bar q\bar q$ system.}
  \label{contributionratio}
\end{figure}

In the following, we examine how the most influential $\langle V_{QQ}^{ce} \rangle$ and $\langle V_{\bar q\bar q}^{cm} \rangle$ depend on the $m_Q/m_{\bar{Q}(\bar{q})}$ ratio in Fig. \ref{contributionratio}. The $ V^{ce} $ provides attraction and repulsion for diquark in color $\bar 3_c$ and $6_c$ configurations, respectively.  For the color $\bar 3_c$ $QQ$,  the attraction from the short-range color-electric potential scales with $\langle \frac{1}{r}\rangle \propto \alpha_s m_{ij}$, where $m_{ij}$ is the reduced mass between the interacting quarks. Then, the color-electric force $V_{QQ}^{ce} $ is more significant. The $ V^{cm} $ depends on  $\frac{1}{m_im_j}\langle \frac{\lambda_i}{2}\frac{\lambda_j}{2} \boldsymbol{\sigma}_i\boldsymbol{\sigma}_j\rangle$, making the $V_{qq}^{cm}$ more significant, while $ V_{Qq}^{cm} $ and $ V_{QQ}^{cm} $ are suppressed by $1/m_Q$.

For $I=1$ $bb\bar q\bar q$ states with $J^{PC}=2^{+}$ and $1^{+-}$, the $QQ$ is in color $\bar 3_c$ configuration. The color-electric interaction $ V_{QQ}^{ce} $ is attractive. As illustrated in Fig.~\ref{contributionratio}, with the increasing $m_Q/m_{\bar{Q}(\bar{q})}$ ratio, the attraction $\langle V_{QQ}^{ce} \rangle $ increases systematically. At the transition from resonance to bound state, the contribution of $\langle V^{ce}_{QQ}\rangle$ changes suddenly. In the $0^{+}$ system, $QQ$ diquark can also be in the $6_c$ representation, generating repulsive color-electric interactions. This leads to less attraction compared to $1^{+}$ and $2^{+}$ states. In the heavy quark limit, the $\bar 3_c$ and $6_c$ diquark will decouple since the spin-flip transition is suppressed as $1/m_Q^2$. Consequently, all the lowest-lying states across three  $J^{PC}$ categories have the color $\bar 3_c$ $QQ$ diquark with spin $S=1$ representation and the same  $ V_{QQ}^{ce} $ attraction due to the heavy quark spin symmetry.

The $I=1$ $bb\bar q\bar q$ consists of color  $3_c$ $\bar q\bar q$ with spin $1$ ($J^P$=$0^+$, $1^+$, or $2^+$) and color $\bar 6_c$ $\bar q\bar q$ with spin $0$ ($J^P$=$0^+$), both providing  repulsive color-magnetic interactions.  Fig. \ref{contributionratio} indicates that the overall repulsive $\langle V_{\bar q\bar q}^{cm} \rangle$ increases gradually as $m_Q/m_{\bar q}$ increases. Despite the slow rise in repulsion (dashed line), the rapidly increasing $QQ$ color-electric attraction (solid line) enhances binding, leading to a bound state at a larger heavy mass than the realistic one in the quark model.

Simultaneously, the $I=0$  $bb\bar q \bar q$  consists of color  $3_c$  $\bar q\bar q$ with spin $0$. The $\langle V_{\bar{q}\bar{q}}^{cm} \rangle$ contributes attractively, aiding in the bound state formation at lower mass ratios.  As discussed in Sec. \ref{sec:rd}, the $I=0$ $0(1^+)$ $bb\bar q \bar q$ states form deep and shallow bound states at the physical ratio $m_Q/m_q=16.6$, while the $I=1$ $bb\bar q \bar q$  do not form bound states. To make it clear, we calculate the contributions of the color-electric interaction $\langle V_{bb}^{ce} \rangle$ and color-magnetic interaction $\langle V_{\bar{q}\bar{q}}^{cm} \rangle$ to both the $I=0$ and $I=1$ $bb\bar q\bar q$ states, and present the results in Table \ref{tab:contribution}.  For the $I=0$  $bb\bar q \bar q$ states, the $bb$ diquark significantly attracts due to the color-electric interaction $\langle V_{bb}^{ce} \rangle$ and
the color-magnetic interaction $\langle V_{qq}^{cm} \rangle$  enhances the binding energy. In contrast, for the $I=1$ states,  despite the color-electric attraction, the $\langle V_{qq}^{cm} \rangle$  is repulsive. The color-electric attraction alone is not sufficient for binding, requiring a larger attractive force from $\langle V_{bb}^{ce} \rangle$ at higher mass ratios.

\section{Summary} \label{sec:summary}
In this study, we have calculated a comprehensive spectrum of the S-wave tetraquark states comprising identical quarks and antiquarks, including the fully heavy tetraquarks $QQ\bar Q'\bar Q'$ and double-heavy tetraquarks $QQ\bar s\bar s$ and $QQ\bar q\bar q$ using the quark model. For the four-body system, the potential is modeled as the sum of all quark-quark and quark-antiquark interactions. We then solve the full four-body Schr{\"o}dinger equation using the Gaussian expansion method and subsequently identify the bound and resonant states through the complex-scaling method. It is crucial to note that in this work, we have assumed the confinement potentials just as the sum of two-quark potentials. Altering this mechanism could lead to different results, such as the four-body string-type confinement potential is shown to yield bound $cc\bar c\bar c$ states and the $X(6600)$ in Ref. \cite{Wang:2023jqs}. These variations in the spectrum could be used as a probe to study the confinement mechanisms in the multi-quark potential.

The spectrum of the tetraquark systems in this work is only the low-lying resonant states primarily up to approximately the scattering thresholds with a ground and a third radially excited states. It is notable that applying the quark model to highly excited states is problematic due to the potential coupled channel effects with other degrees of freedom, such as the interactions with hadron loops. This challenges its reliability for describing such states. Furthermore,  the limitations of the CSM method and the exclusion of the scattering states with P-wave excitations in the higher-lying resonances may challenge their identification and render the results less reliable as discussed in the literature.  Nevertheless, for the low-lying resonant states, which dominantly interact with S-wave scattering states, the quark model provides more trustworthy insights into the quark-quark interactions, making the results more reliable. We summarize our results as follows: 

\begin{itemize}
  
\item  For the S-wave full-charmed tetraquark state $cc\bar c\bar c$, we do not find a bound state but identify  $6 $ resonances: each $2$ for $J^{PC}=0^{++}$, $1^{+-}$ and $2^{++}$. The lower and higher ones are located approximately at $E(\Gamma)=7000(70)$ MeV and $7200(30)$ MeV, with uncertainties up to $20$ MeV due to CSM. The mass discrepancy between the AL1 and BGS quark models is about up to $60$ MeV. Compared with the experiment, the lower one is at $100$ MeV higher than $X(6900)$ while the higher resonance is consistent with the observed $X(7200)$. The experimental $X(6600)$ is not detected using our current model.

\item 
The $bb\bar b\bar b$ resonances have a similar resonance pattern as $cc\bar c\bar c$, and no bound states are found. We have identified $6$ resonances at around $E(\Gamma)=19790(60)$ and $19960 (30)$ MeV. No $bb\bar b\bar b$ states have been observed yet. Further exploration of the  $bb\bar b\bar b$ resonances in LHCb and CMS collaborations is highly desirable.  For the $bb\bar c\bar c$, we found $10$ resonances. Within each $J^P$ quantum number category, there are three resonances distributed around $13450$, $13570$, and $13680$ MeV. Additionally, a $2^{++}$ state is identified around $13652$ MeV.

 \item For the $QQ\bar s\bar s$ and the $I=1$ $QQ\bar q\bar q$ systems, we have identified $35$ resonances without any bound states.  Notably,  there are  $bb\bar s\bar s$ and $I=1$  $bb\bar q\bar q$ states located closely to their respective lowest scattering states, as permitted by their $J^{P}$ quantum number. However, the charmed analogs of such states are absent. This is attributed to the weaker short-range color-Coulomb attractions, which result in less compact $cc\bar s\bar s$ and $I=1$ $cc\bar q\bar q$ states,  and stronger coupling to $D^{(*)}D^{(*)}$ scattering states. 

 \item Within our framework, we only find the $I=0$ $QQ\bar q\bar q$ system with $I(J^P)=0(1^+)$ is possible to form the bound states. We identified a bound $cc\bar q\bar q$ state with a binding energy -$15$ MeV,  which contrasts markedly with the near-threshold $T_{cc}$ states with a minimal binding energy of  $-361 \pm 40$ keV. Additionally, there exists a deeply bound $bb\bar q\bar q$ with the binding energy $-153$ MeV, which can be only explored through the weak decays. The existence is consistent with Lattice QCD calculations albeit a wide distribution of the binding energies from around $-60 $ MeV to $-190$ MeV. Another shallow bound state is, $-4$ MeV below the $BB^*$ threshold. One should note that near-threshold states may need to consider the additional contributions from the long-range interactions as discussed in the literature. 

 \item  In a $QQ\bar q\bar q$ system, the color-electric forces contribute attractively for the color $\bar 3_c$ $QQ$ and the attractions keep increasing as the mass ratio $m_Q/m_q$ increases. A sufficiently large ratio can lead to the bound state. The color-magnetic interactions are attractive for the $\bar 3_c$ $qq$ with spin $0$, which combine with the color-electric force to enable the bound state formation at realistic mass ratios. This mechanism underlies why the $I=0$ $QQ\bar q\bar q$ ($Q=b,c$) form the bound state, while the $I=1$ counterparts cannot at a realistic mass ratio. 
\end{itemize}

The majority of the predicted tetraquark spectrum has not been explored in experiments.  Investigating these states in LHCb, CMS, ATLAS, or other ongoing or planned collaborations will have a significant impact on the understanding of tetraquark spectroscopy. Additionally, it will help to probe the enigmatic confinement mechanisms and inner structures within the four-body system.

\begin{acknowledgements}
 Q.~Meng is supported by the National Natural Science Foundation of China (Grant No. 12275129), the Fundamental Research Funds for the Central Universities (Grant No. 020414380209), and the Jiangsu Funding Program for Excellent Postdoctoral Talent (Grant No. 2022ZB9).
 G.J.~Wang is 
supported by JSPS KAKENHI (No. 23K03427). M.~Oka is supported in part by the
JSPS KAKENHI (Nos.~19H05159, 20K03959, 21H00132 and 23K03427). 
\end{acknowledgements}

\begin{appendix}
\section{The mass spectra of the heavy mesons $Q\bar q$ and heavy quarkonium $Q\bar Q'$} \label{appendix-mm}
We present the mass spectra for ground and excited states of the heavy quarkonium ($Q\bar Q'$) in Table \ref{tab:hq}  and heavy mesons ($Q\bar q$) in Table \ref{tab:hm}, respectively. The results are extended up to the third radially excited states. In our tetraquark calculations, we focus on the lower resonances, which are located up to near the threshold containing a third radially excited meson.  It should be noted that the convergence of the excited resonances, particularly those above the thresholds of scattering states involving higher excited states, are not good anymore and thus less reliable. Moreover, the scattering states with higher excited orbital heavy mesons or quarkonium may also contribute more importantly, leading to the less reliable highly excited resonances. 
\begin{center}
\renewcommand\arraystretch{1.0} 
\begin{table}[t]
\caption{The mass spectra of heavy quarkonium $Q\bar Q'$ (in units of MeV), obtained using the AL1 potential (AL1) \cite{Silvestre-Brac:1996myf} are presented alongside with experimental values (Expt.) \cite{zyla2020prog} for a comparison. Additionally, the masses of charmonium calculated using the BGS potential (BGS) \cite{Barnes:2005pb} are also listed.}
\label{tab:hq}
\begin{tabular}{p{1.5cm}<{\centering}p{1.5cm}<{\centering} p{1.8cm}<{\centering} p{2.2cm}<{\centering} p{1.5cm}<{\centering} p{1.5cm}<{\centering} }
\hline\hline
 meson      & $J^P$ & state & Expt. & AL1 & BGS \\
\hline
 $c\bar{c}$ & $0^-$ & $1S$ & 2984 & 3005 & 2982 \\
            & $0^-$ & $2S$ & 3638 & 3608 & 3630 \\
            & $0^-$ & $3S$ &      & 4027 & 4047 \\
\cline{2-6}
            & $1^-$ & $1S$ & 3097 & 3101 & 3090 \\
            & $1^-$ & $2S$ & 3686 & 3641 & 3672 \\
            & $1^-$ & $3S$ & 4039 & 4049 & 4074 \\
\hline
 $b\bar{b}$ & $0^-$ & $1S$ & 9399 & 9424 &  \\
            & $0^-$ & $2S$ & 9999 & 10003 &  \\
            & $0^-$ & $3S$ &      & 10335 &  \\
\cline{2-6}
            & $1^-$ & $1S$ & 9460  & 9462 &  \\
            & $1^-$ & $2S$ & 10023 & 10013 &  \\
            & $1^-$ & $3S$ & 10351 & 10341 &  \\
\hline
 $b\bar{c}$ & $0^-$ & $1S$ & 6274 & 6292 &  \\
            & $0^-$ & $2S$ & 6871 & 6855 &  \\
            & $0^-$ & $3S$ &      & 7226 &  \\
\cline{2-6}
            & $1^-$ & $1S$ &      & 6343 &  \\
            & $1^-$ & $2S$ &      & 6871 &  \\
            & $1^-$ & $3S$ &      & 7238 &  \\

\hline\hline
\end{tabular}
\end{table}
\end{center}

\begin{center}
\renewcommand\arraystretch{1.0} 
\begin{table}[t]
\caption{The mass spectra of heavy mesons $Q\bar q$ (in units of MeV), obtained using the AL1 potential (AL1) \cite{Silvestre-Brac:1996myf} are presented, alongside with experimental values (Expt.) \cite{zyla2020prog} for a comparison.}
\label{tab:hm}
\begin{tabular}{p{1.5cm}<{\centering}p{1.5cm}<{\centering} p{1.8cm}<{\centering} p{2.2cm}<{\centering} p{1.5cm}<{\centering} p{1.5cm}<{\centering} }
\hline\hline
 meson      & $J^P$ & state & Expt. & AL1 & BGS \\
\hline
 $c\bar{s}$ & $0^-$ & $1S$ & 1968 & 1962 &  \\
            & $0^-$ & $2S$ &  & 2663 &  \\
            & $0^-$ & $3S$ &  & 3169 &  \\
\cline{2-6}
            & $1^-$ & $1S$ & 2112 & 2102 &  \\
            & $1^-$ & $2S$ &  & 2721 &  \\
            & $1^-$ & $3S$ &  & 3215 &  \\
\hline
 $b\bar{q}$ & $0^-$ & $1S$ & 5279 & 5293 &  \\
            & $0^-$ & $2S$ &  & 6013 &  \\
            & $0^-$ & $3S$ &  & 6586 &  \\
\cline{2-6}
            & $1^-$ & $1S$ & 5325 & 5351 &  \\
            & $1^-$ & $2S$ &  & 6040 &  \\
            & $1^-$ & $3S$ &  & 6606 &  \\
\hline
 $c\bar{q}$ & $0^-$ & $1S$ & 1870 & 1862 &  \\
            & $0^-$ & $2S$ & 2549 & 2643 &  \\
            & $0^-$ & $3S$ &  & 3238 &  \\
\cline{2-6}
            & $1^-$ & $1S$ & 2010 & 2016 &  \\
            & $1^-$ & $2S$ &  & 2715 &  \\
            & $1^-$ & $3S$ &  & 3287 &  \\
 \hline
 $b\bar{s}$ & $0^-$ & $1S$ & 5367 & 5361 &  \\
            & $0^-$ & $2S$ &  & 5999 &  \\
            & $0^-$ & $3S$ &  & 6467 &  \\
\cline{2-6}
            & $1^-$ & $1S$ & 5415 & 5417 &  \\
            & $1^-$ & $2S$ &  & 6023 &  \\
            & $1^-$ & $3S$ &  & 6485 &  \\
\hline\hline
\end{tabular}
\end{table}
\end{center}

\section{The eigenvalues in CSM} \label{sec:ecsm}
\begin{figure*}
\centering \subfigure[]{
\begin{minipage}[t]{0.35\linewidth}
\includegraphics[width=1\textwidth]{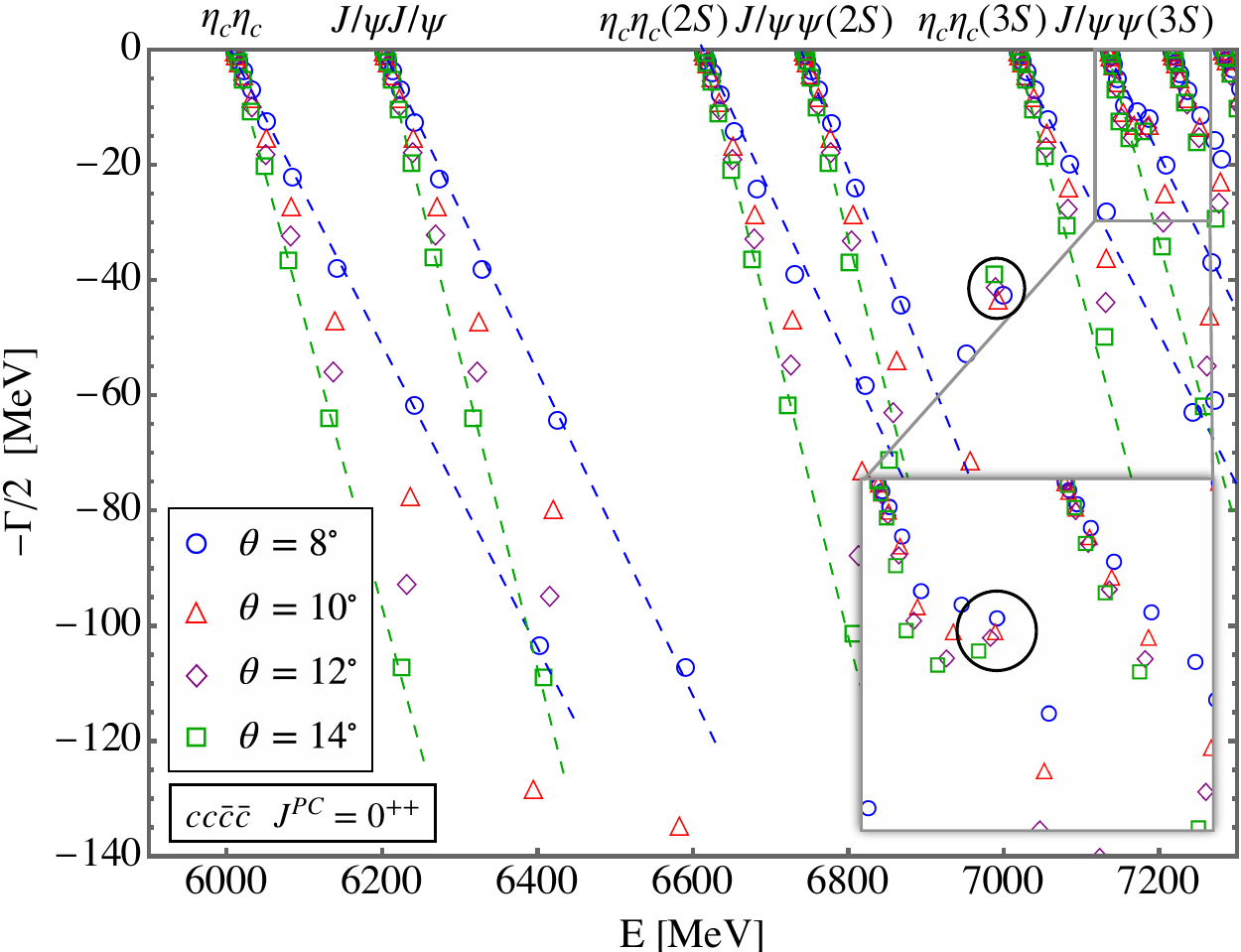}
 \label{fig:cccc0}
\end{minipage}%
}%
\subfigure[]{
\begin{minipage}[t]{0.35\linewidth}
\centering
\includegraphics[width=1\textwidth]{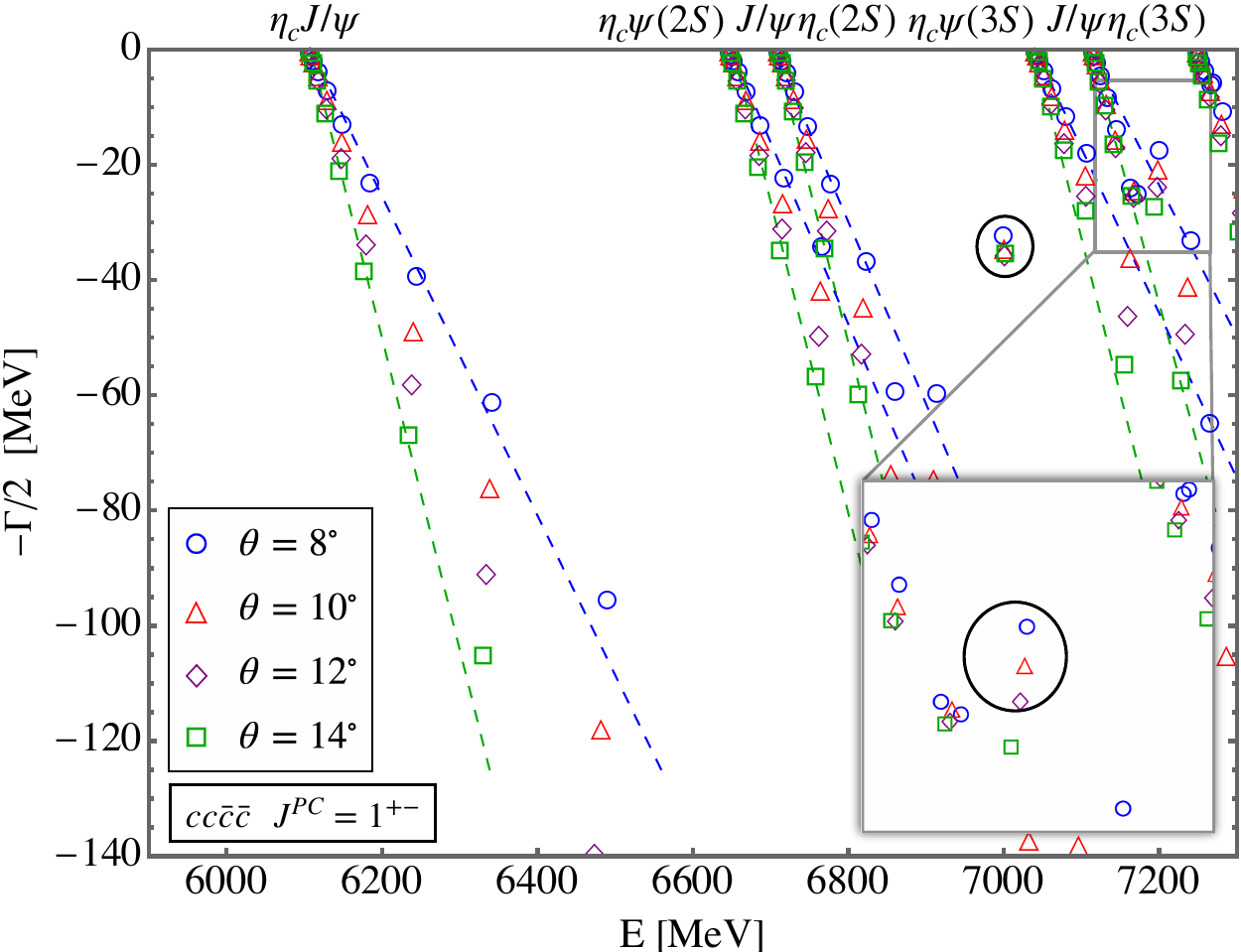}     \label{fig:cccc1}

\end{minipage}%
}%
\subfigure[]{
\begin{minipage}[t]{0.35\linewidth}
\centering
\includegraphics[width=1\textwidth]{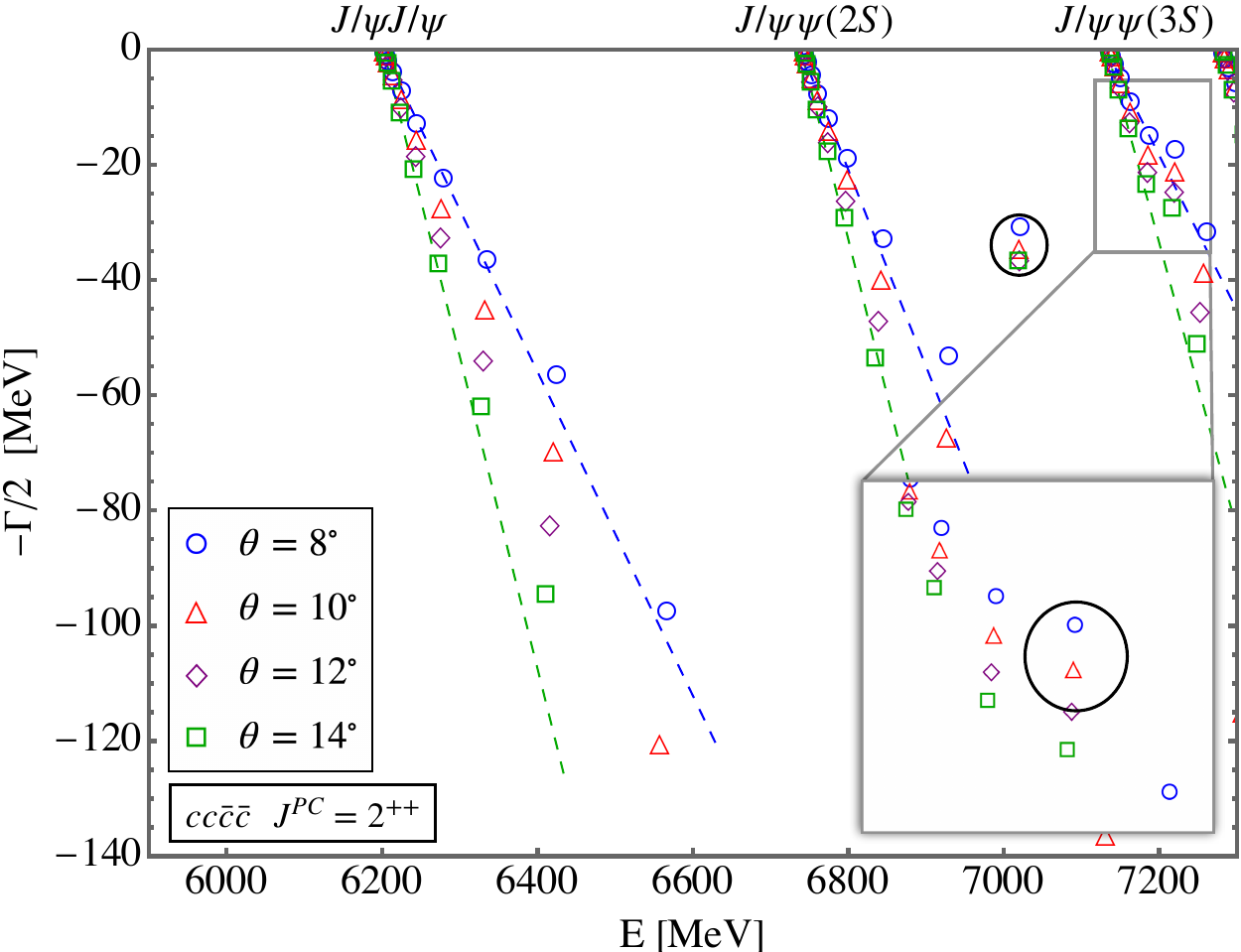}
    \label{fig:cccc2}
\end{minipage}%

}%
\caption{ The complex eigenvalues of the ${cc\bar c\bar c}$ states with $J^{PC}$ quantum numbers (a) $0^{++}$, (b) $1^{+-}$, and (c) $2^{++}$.  The results are obtained with varying $\theta$ in the complex-scaling method. The inset provides a detailed view of the selected energy region, highlighting the resonances marked in black circles. } \label{fig:cccc}
\end{figure*}

\begin{figure*}
\centering \subfigure[]{
\begin{minipage}[t]{0.35\linewidth}
\includegraphics[width=1\textwidth]{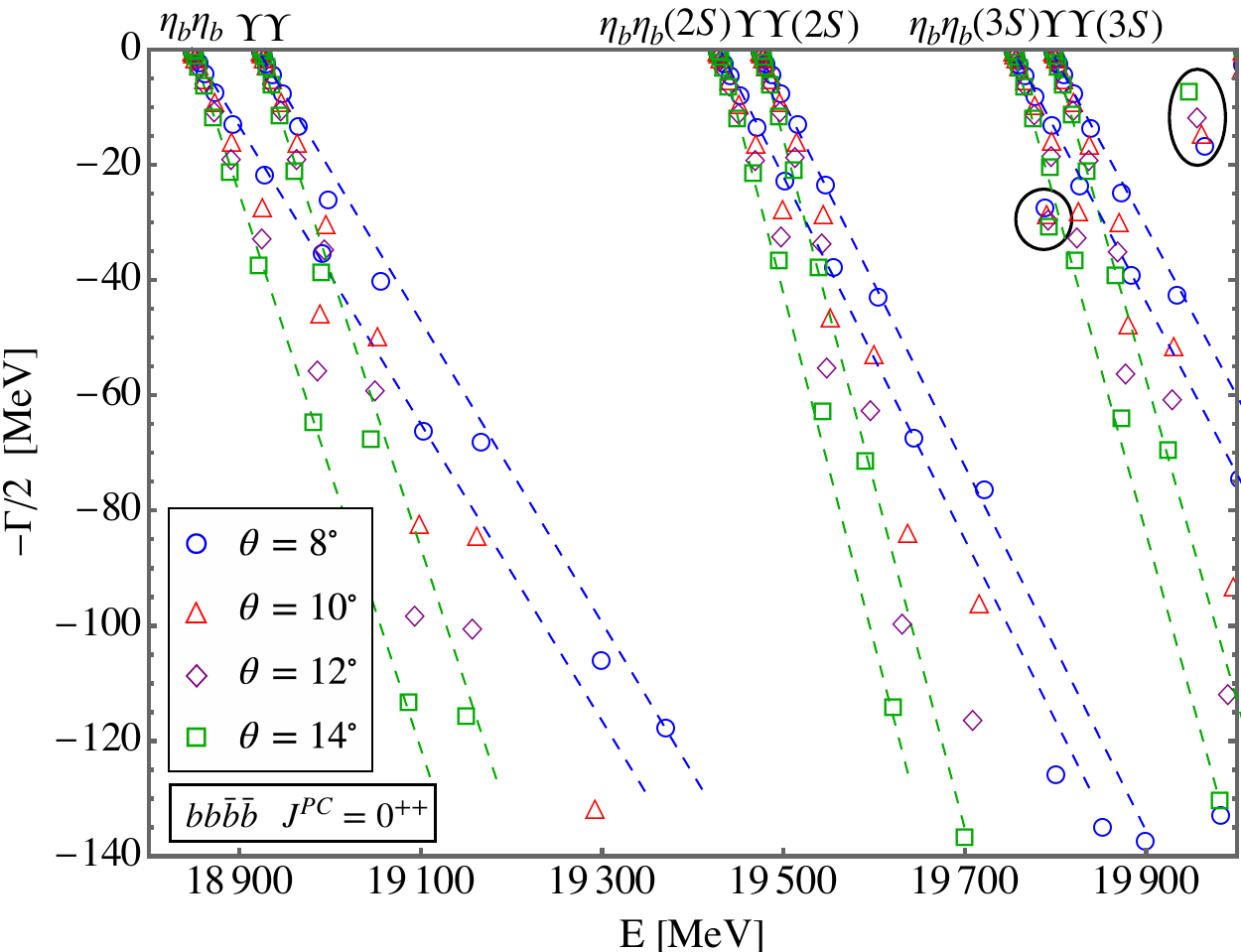}
 \label{fig:bbbb0}
\end{minipage}%
}%
\subfigure[]{
\begin{minipage}[t]{0.35\linewidth}
\centering
\includegraphics[width=1\textwidth]{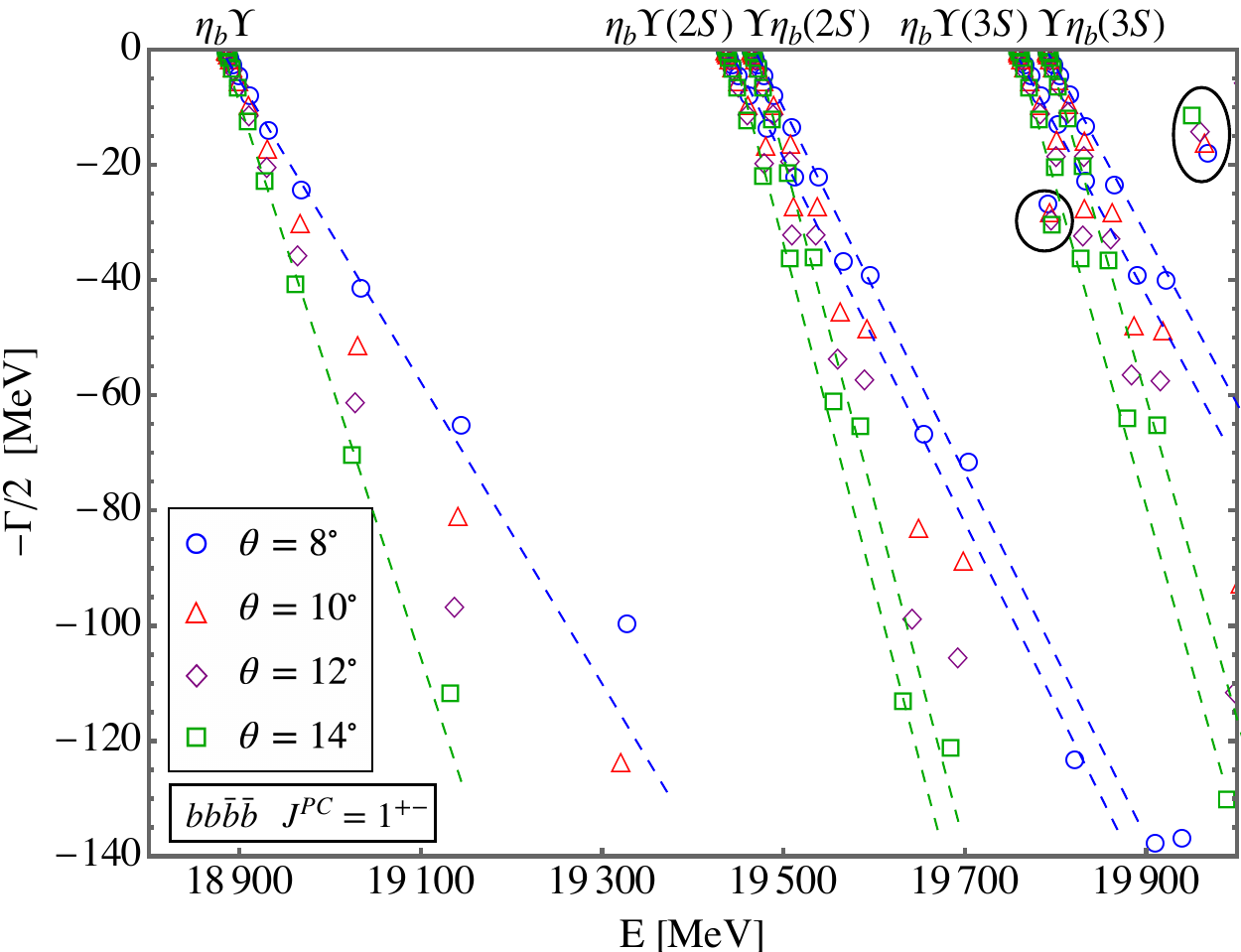}     \label{fig:bbbb1}

\end{minipage}%
}%
\subfigure[]{
\begin{minipage}[t]{0.35\linewidth}
\centering
\includegraphics[width=1\textwidth]{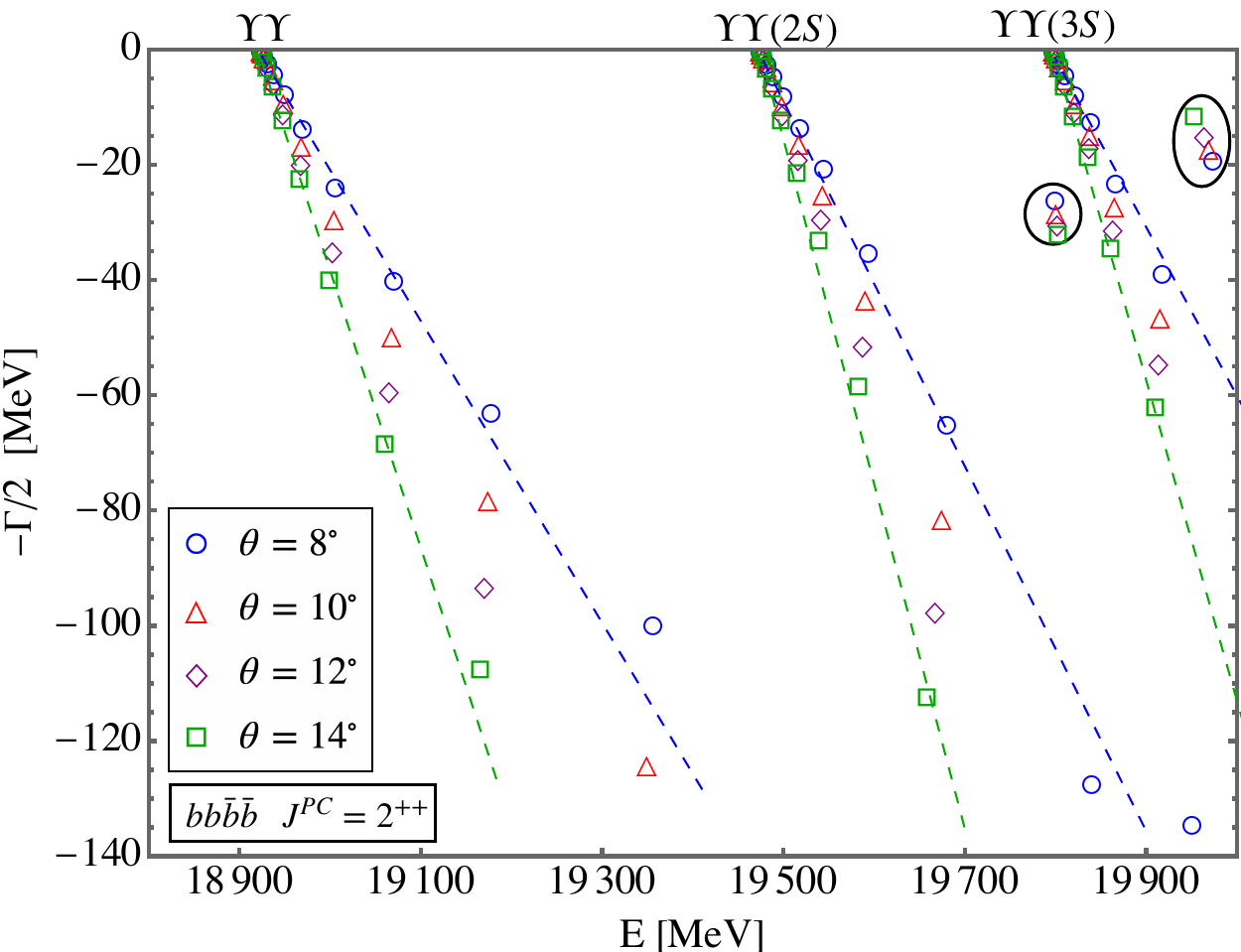}
    \label{fig:bbbb2}
\end{minipage}%

}%
\caption{ The complex eigenvalues of the ${bb\bar b\bar b}$ states with $J^{PC}$ quantum numbers (a) $0^{++}$, (b) $1^{+-}$, and (c) $2^{++}$. The results are obtained with varying $\theta$ in the complex-scaling method. The inset provides a detailed view of the selected energy region, highlighting the resonances marked in black circles.} \label{fig:bbbb}
\end{figure*}

\begin{figure*}
\centering \subfigure[]{
\begin{minipage}[t]{0.35\linewidth}
\includegraphics[width=1\textwidth]{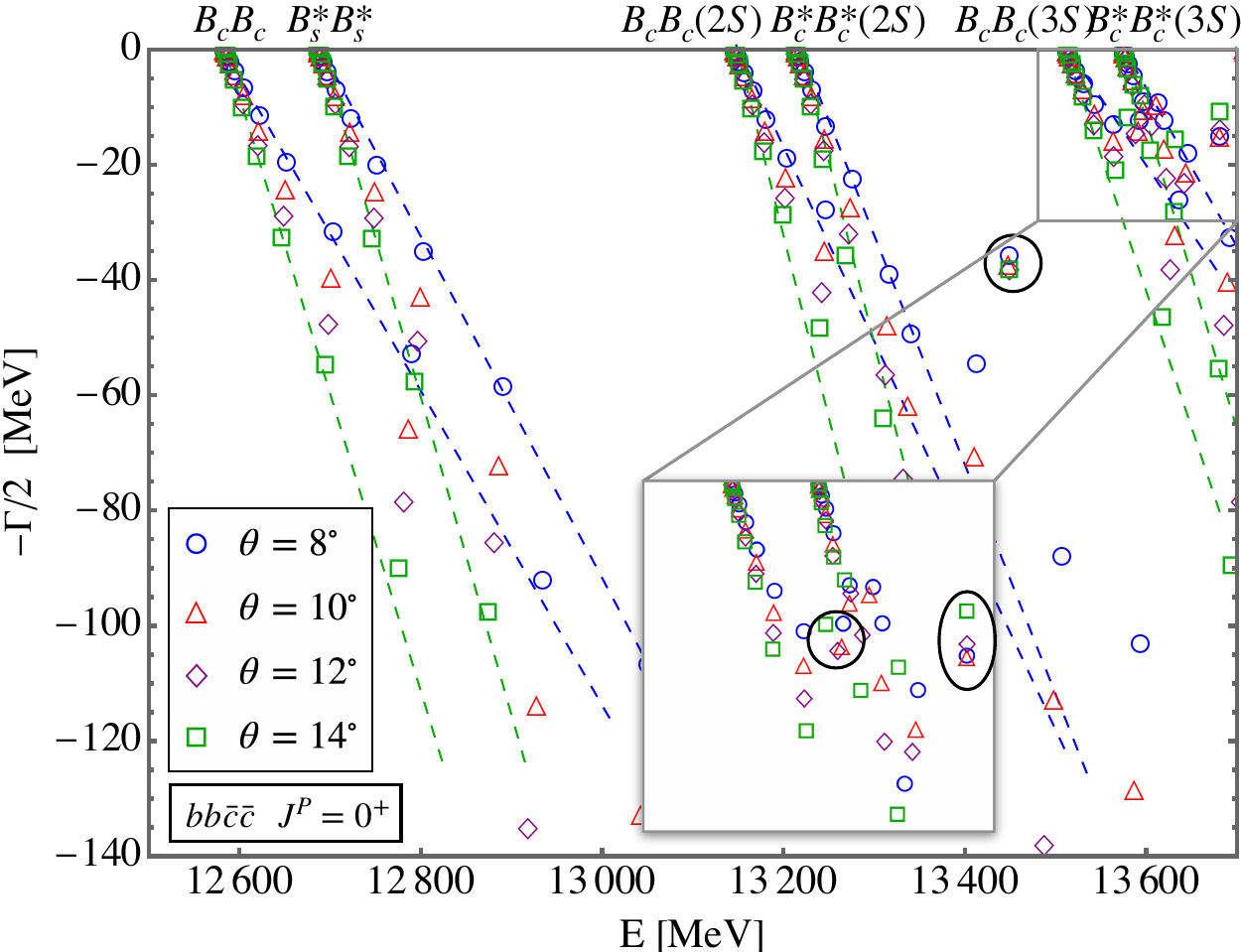}
 \label{fig:bbcc0}
\end{minipage}%
}%
\subfigure[]{
\begin{minipage}[t]{0.35\linewidth}
\centering
\includegraphics[width=1\textwidth]{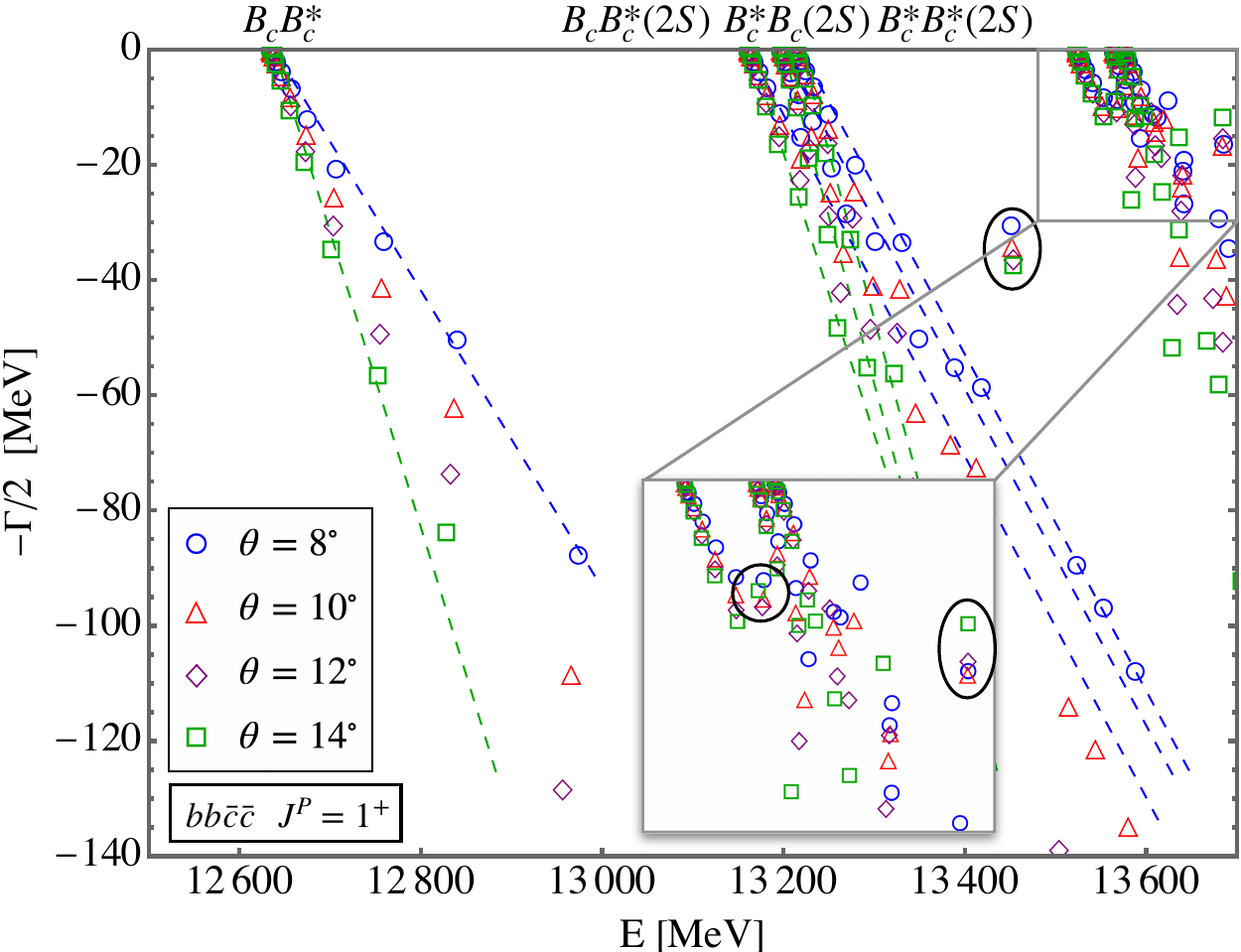}     \label{fig:bbcc1}

\end{minipage}%
}%
\subfigure[]{
\begin{minipage}[t]{0.35\linewidth}
\centering
\includegraphics[width=1\textwidth]{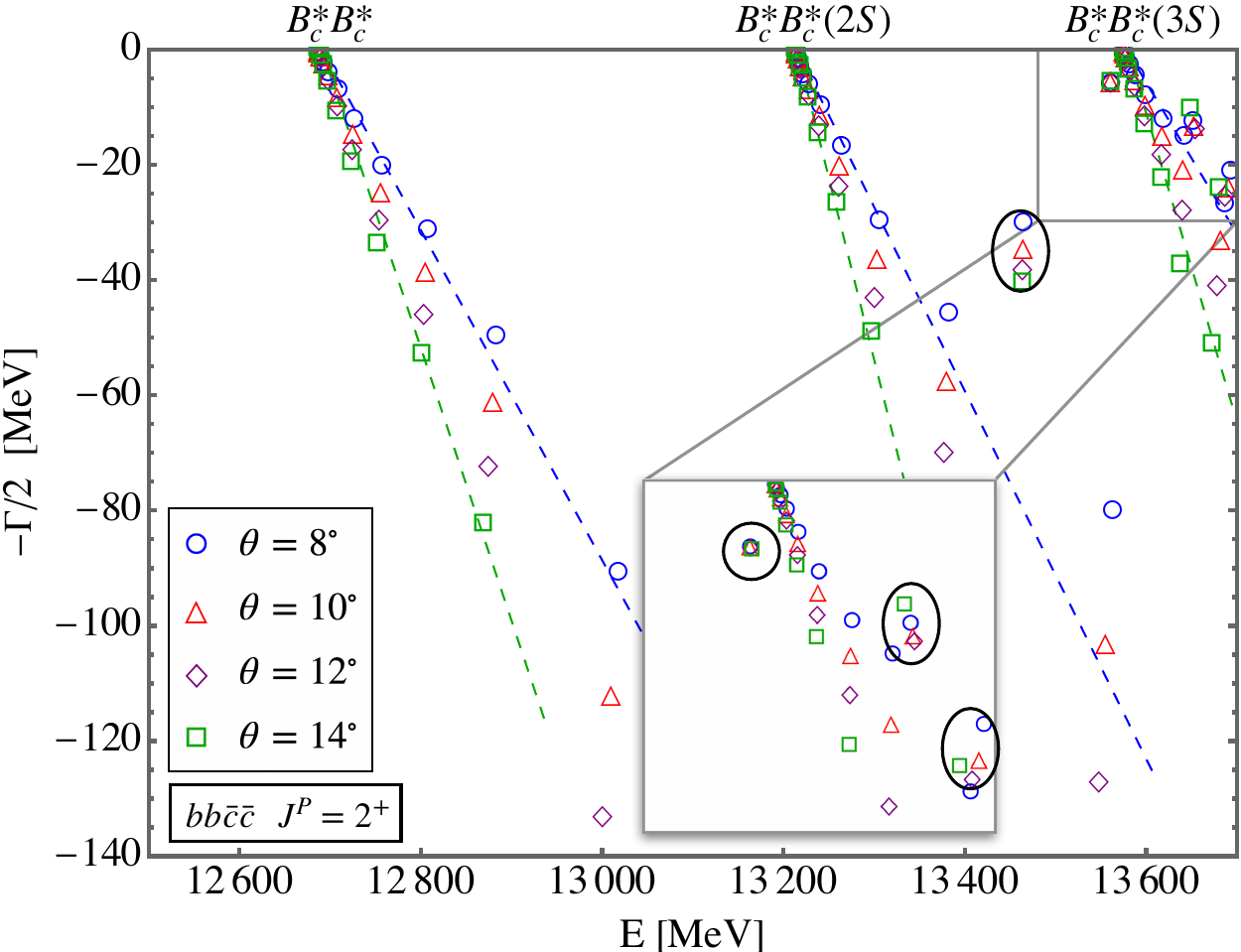}
    \label{fig:bbcc2}
\end{minipage}%

}%
\caption{ The complex eigenvalues of the ${bb\bar c\bar c}$ ($cc\bar b\bar b$) states with $J^{PC}$ quantum numbers (a) $0^{+}$, (b) $1^{+}$, and (c) $2^{+}$. The results are obtained with varying $\theta$ in the complex-scaling method. The inset provides a detailed view of the selected energy region, highlighting the resonances marked in black circles.} \label{fig:bbcc}
\end{figure*}

\begin{figure*}
\centering \subfigure[]{
\begin{minipage}[t]{0.35\linewidth}
\includegraphics[width=1\textwidth]{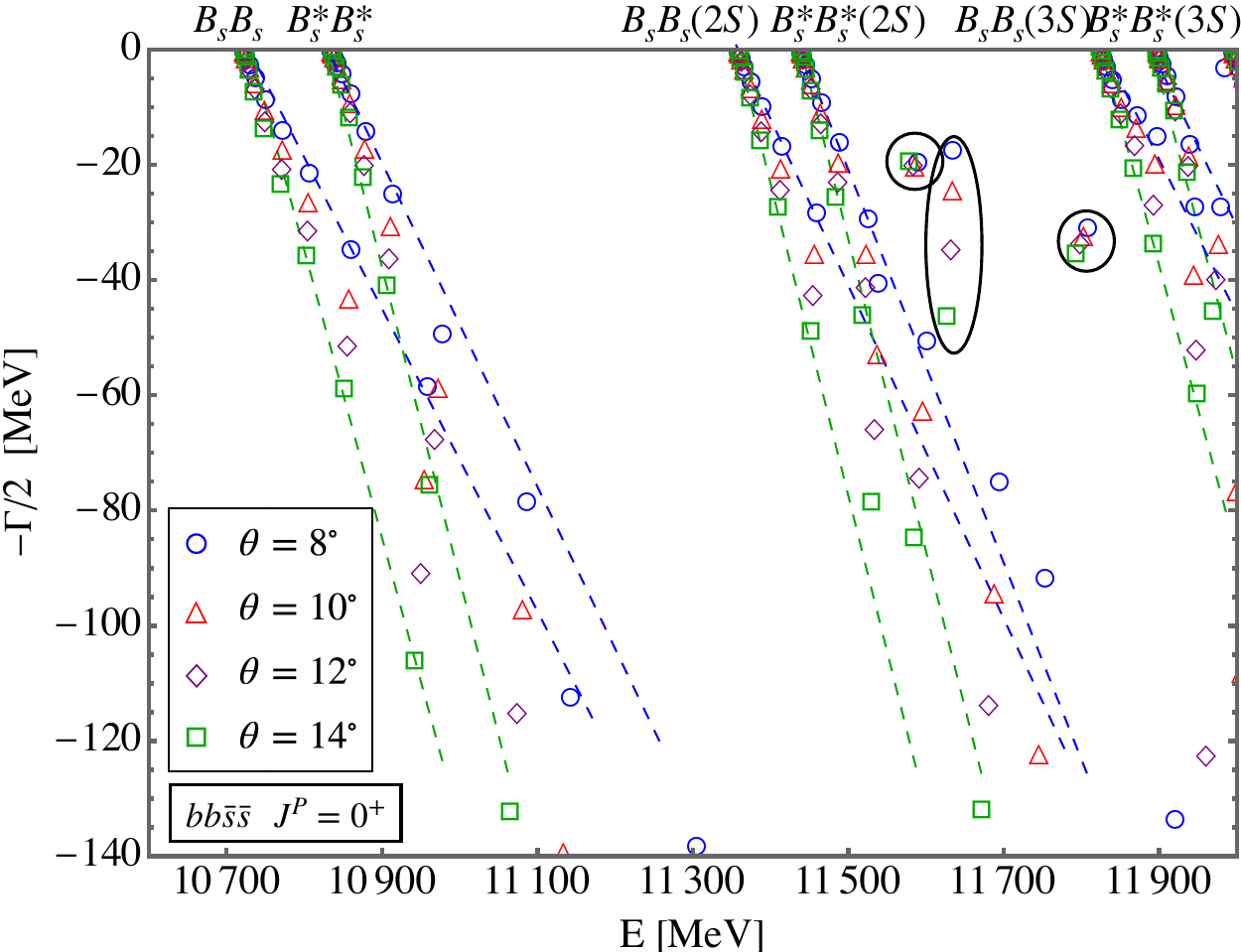}
 \label{fig:bbss0}
\end{minipage}%
}%
\subfigure[]{
\begin{minipage}[t]{0.35\linewidth}
\centering
\includegraphics[width=1\textwidth]{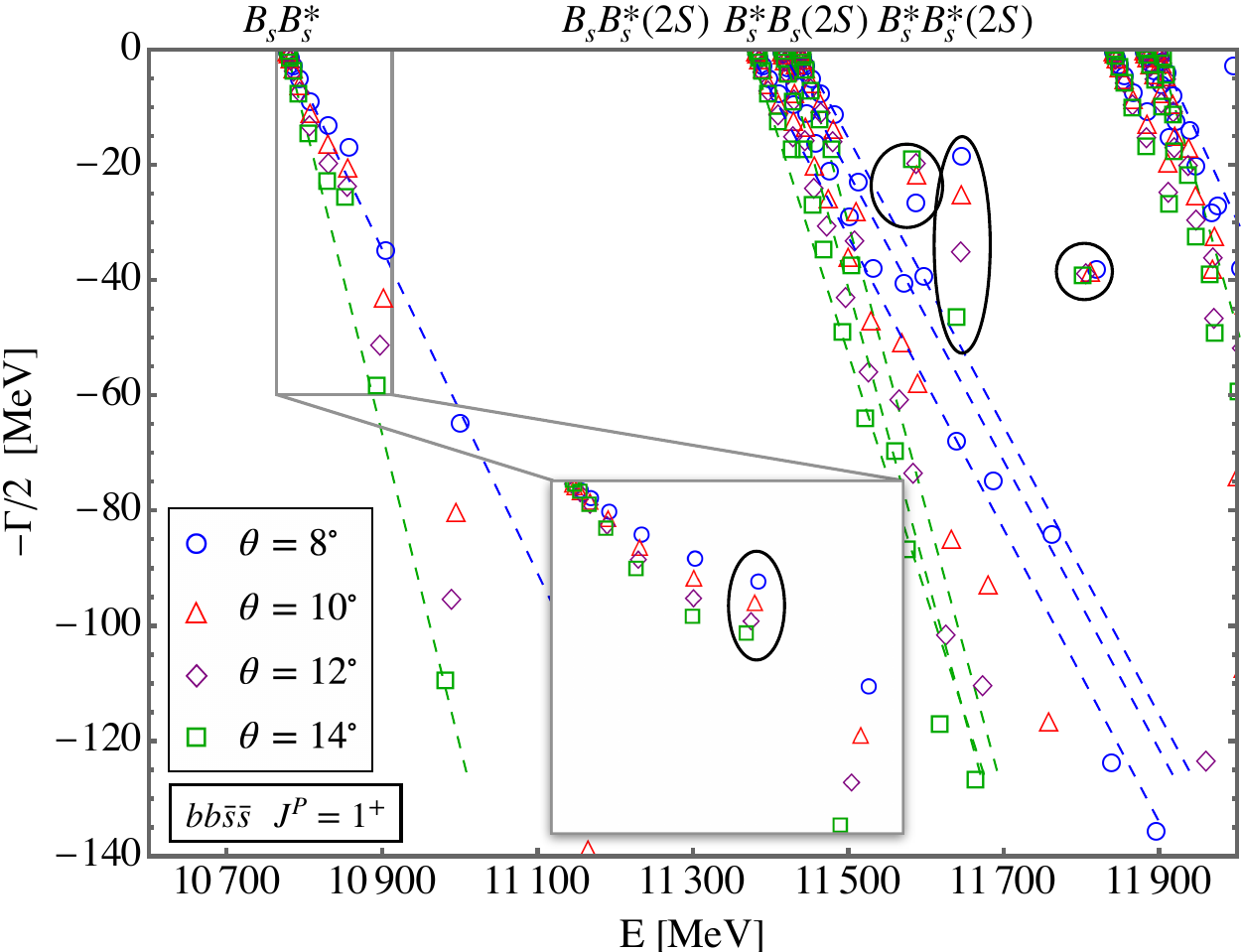}     \label{fig:bbss1}

\end{minipage}%
}%
\subfigure[]{
\begin{minipage}[t]{0.35\linewidth}
\centering
\includegraphics[width=1\textwidth]{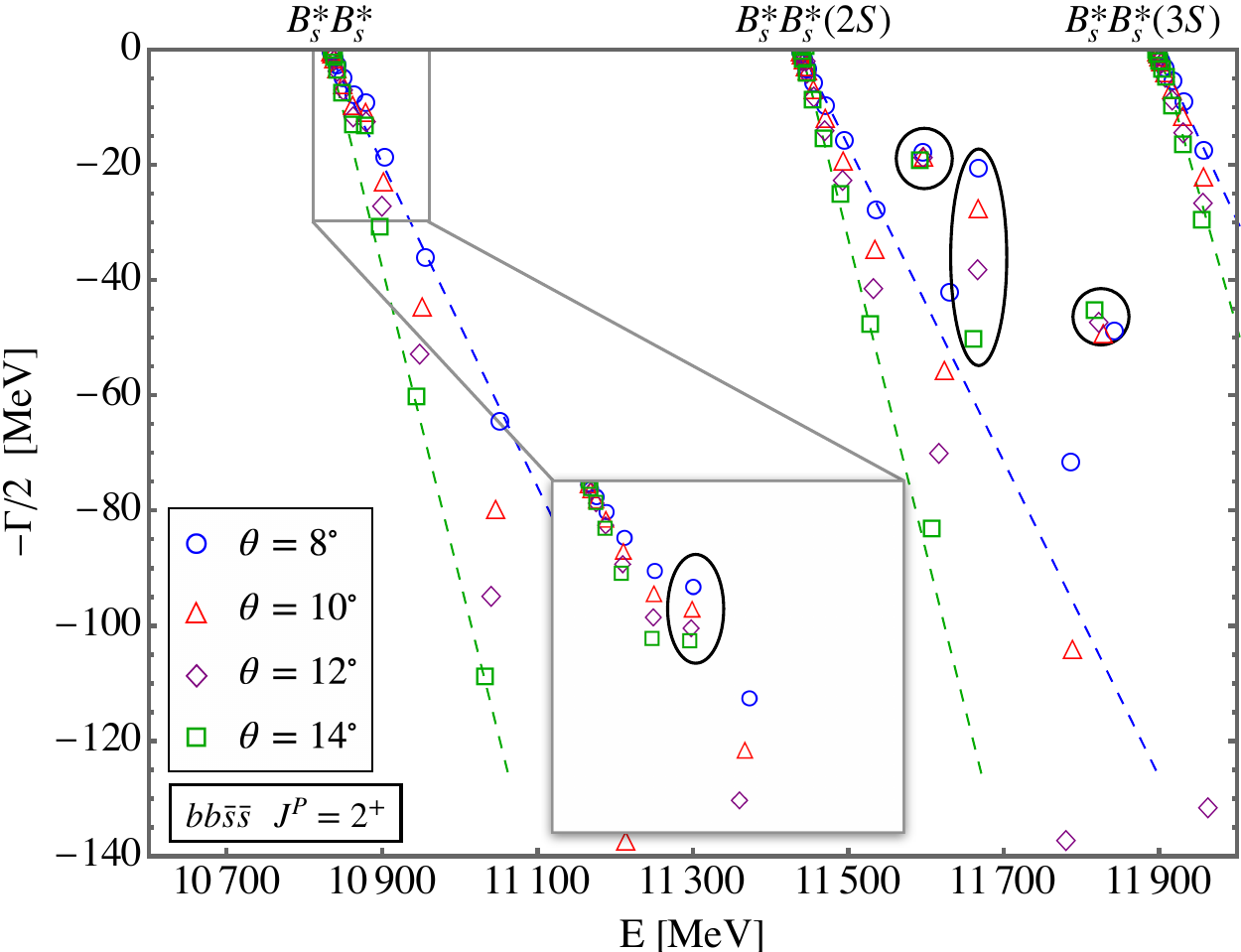}
    \label{fig:bbbb2}
\end{minipage}%

}%
\caption{ The complex eigenvalues of the ${bb\bar s\bar s}$ ($ss\bar b\bar b$) states with $J^{PC}$ quantum numbers (a) $0^{+}$, (b) $1^{+}$, and (c) $2^{+}$. The results are obtained with varying $\theta$ in the complex-scaling method. The inset provides a detailed view of the selected energy region, highlighting the resonances marked in black circles.} \label{fig:bbss}
\end{figure*}

\begin{figure*}
\centering \subfigure[]{
\begin{minipage}[t]{0.35\linewidth}
\includegraphics[width=1\textwidth]{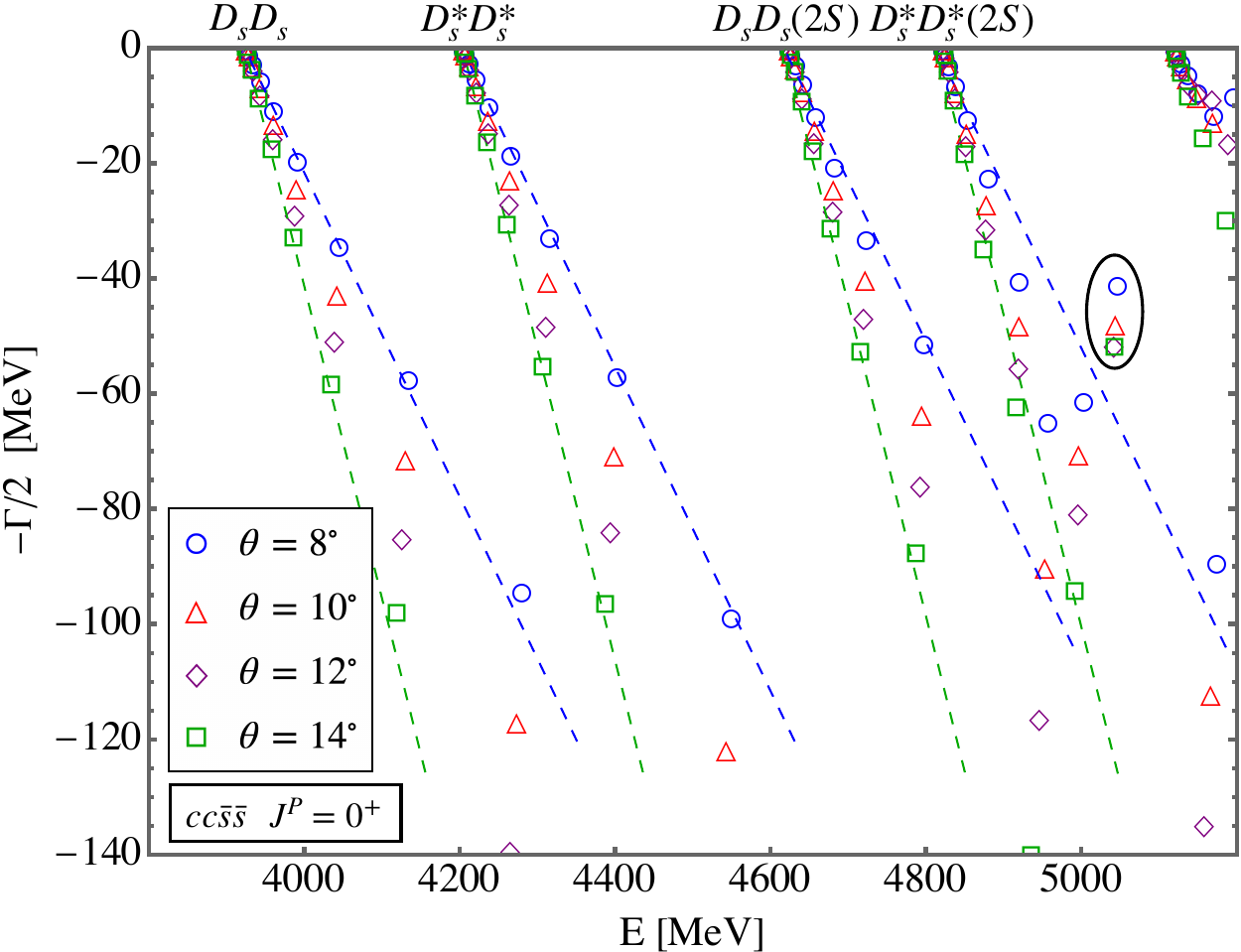}
 \label{fig:ccss0}
\end{minipage}%
}%
\subfigure[]{
\begin{minipage}[t]{0.35\linewidth}
\centering
\includegraphics[width=1\textwidth]{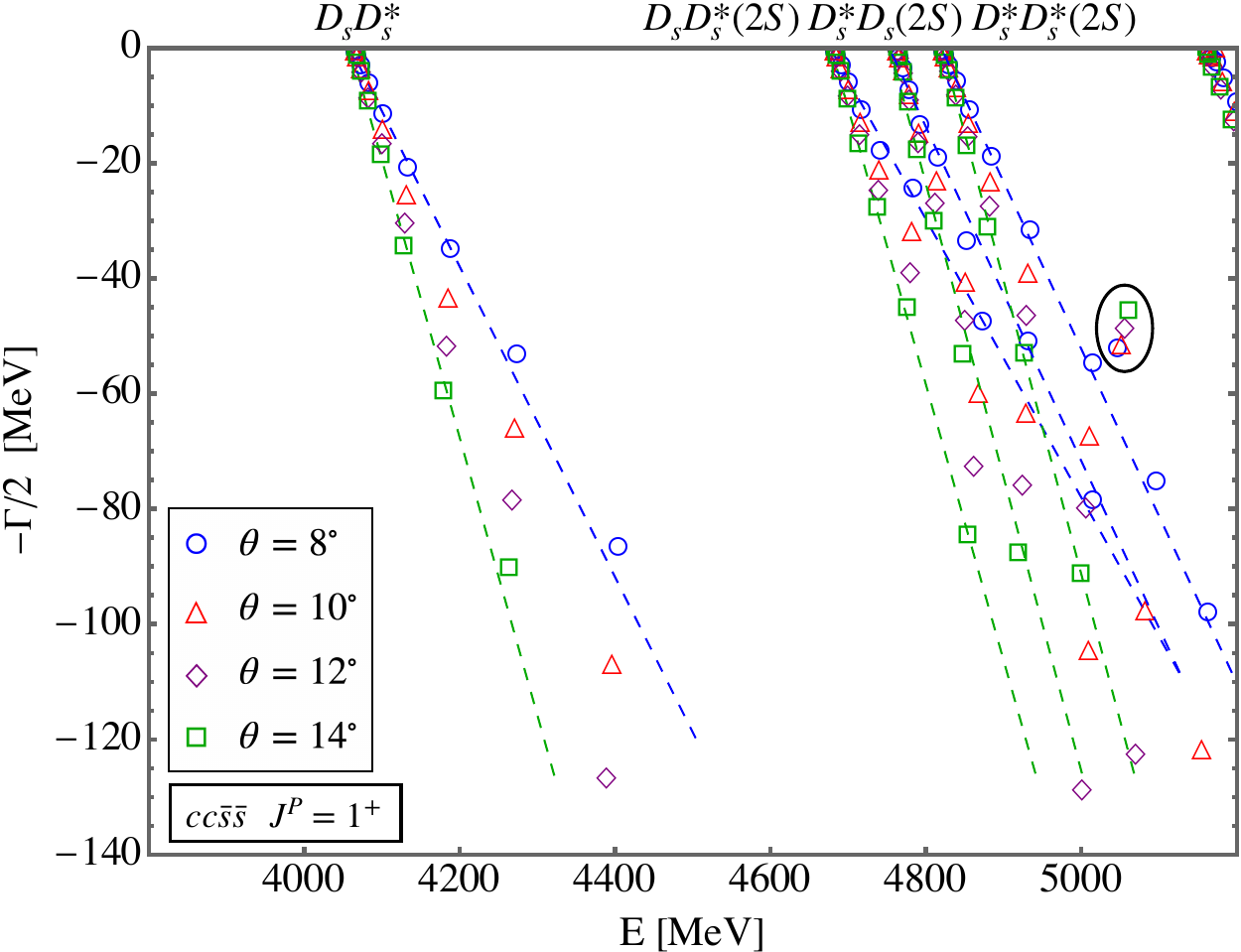}     \label{fig:ccss1}

\end{minipage}%
}%
\subfigure[]{
\begin{minipage}[t]{0.35\linewidth}
\centering
\includegraphics[width=1\textwidth]{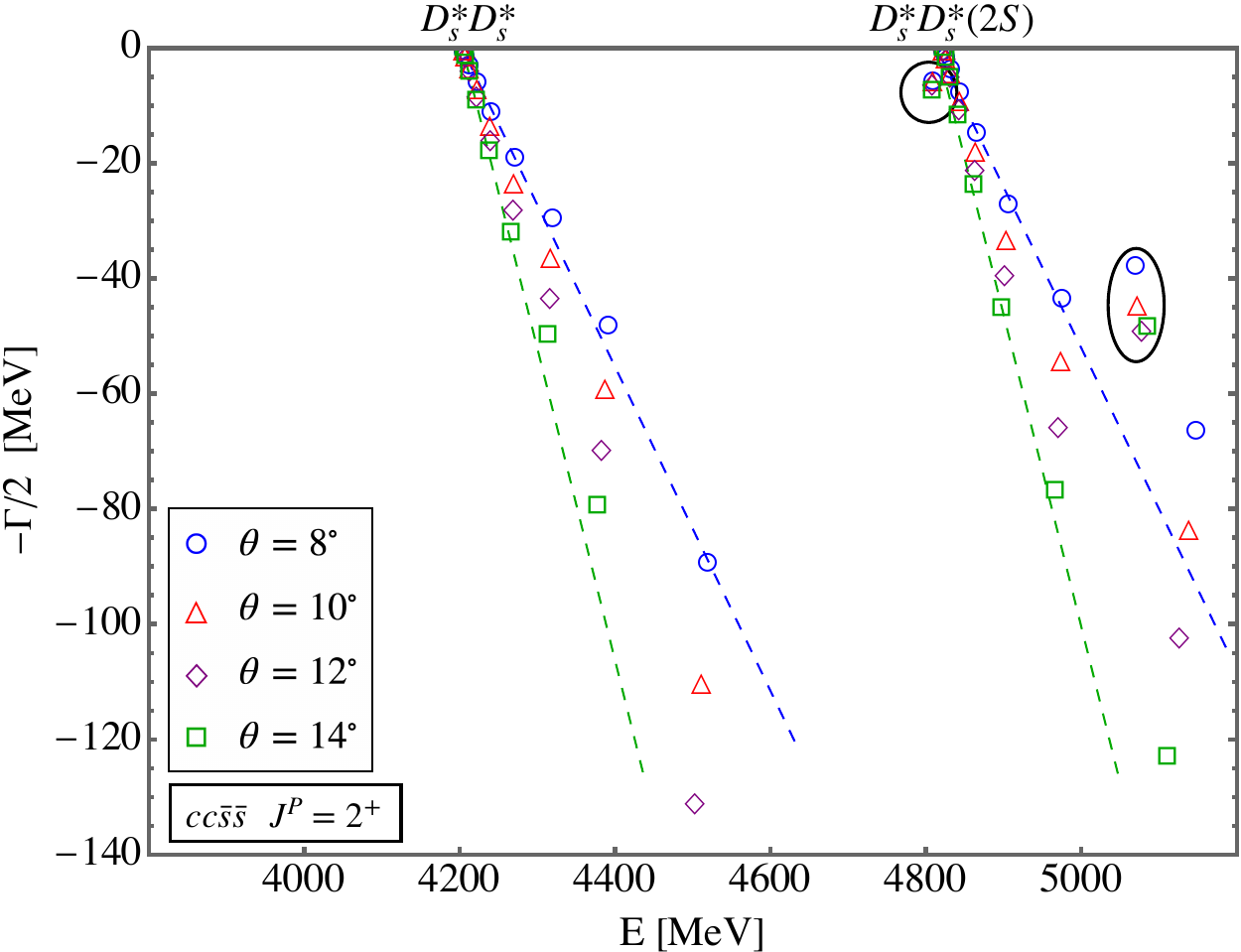}
    \label{fig:ccss2}
\end{minipage}%

}%
\caption{ The complex eigenvalues of the ${cc\bar s\bar s}$ ($ss\bar c\bar c$) states with $J^{PC}$ quantum numbers (a) $0^{+}$, (b) $1^{+}$, and (c) $2^{+}$. The results are obtained with varying $\theta$ in the complex-scaling method. The inset provides a detailed view of the selected energy region, highlighting the resonances marked in black circles.} \label{fig:ccss}
\end{figure*}

\begin{figure*}
\centering \subfigure[]{
\begin{minipage}[t]{0.35\linewidth}
\includegraphics[width=1\textwidth]{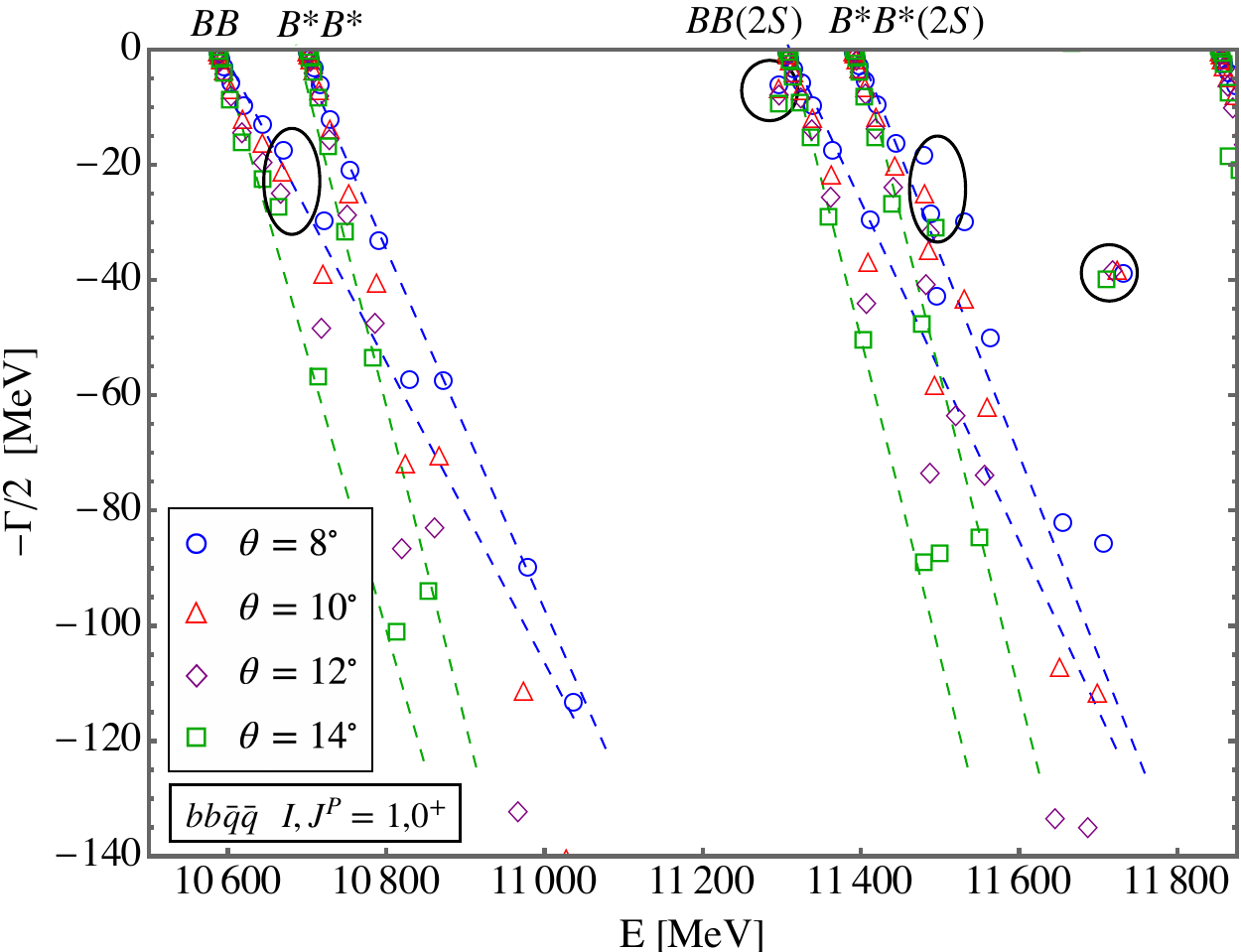}
 \label{fig:bbqq0}
\end{minipage}%
}%
\subfigure[]{
\begin{minipage}[t]{0.35\linewidth}
\centering
\includegraphics[width=1\textwidth]{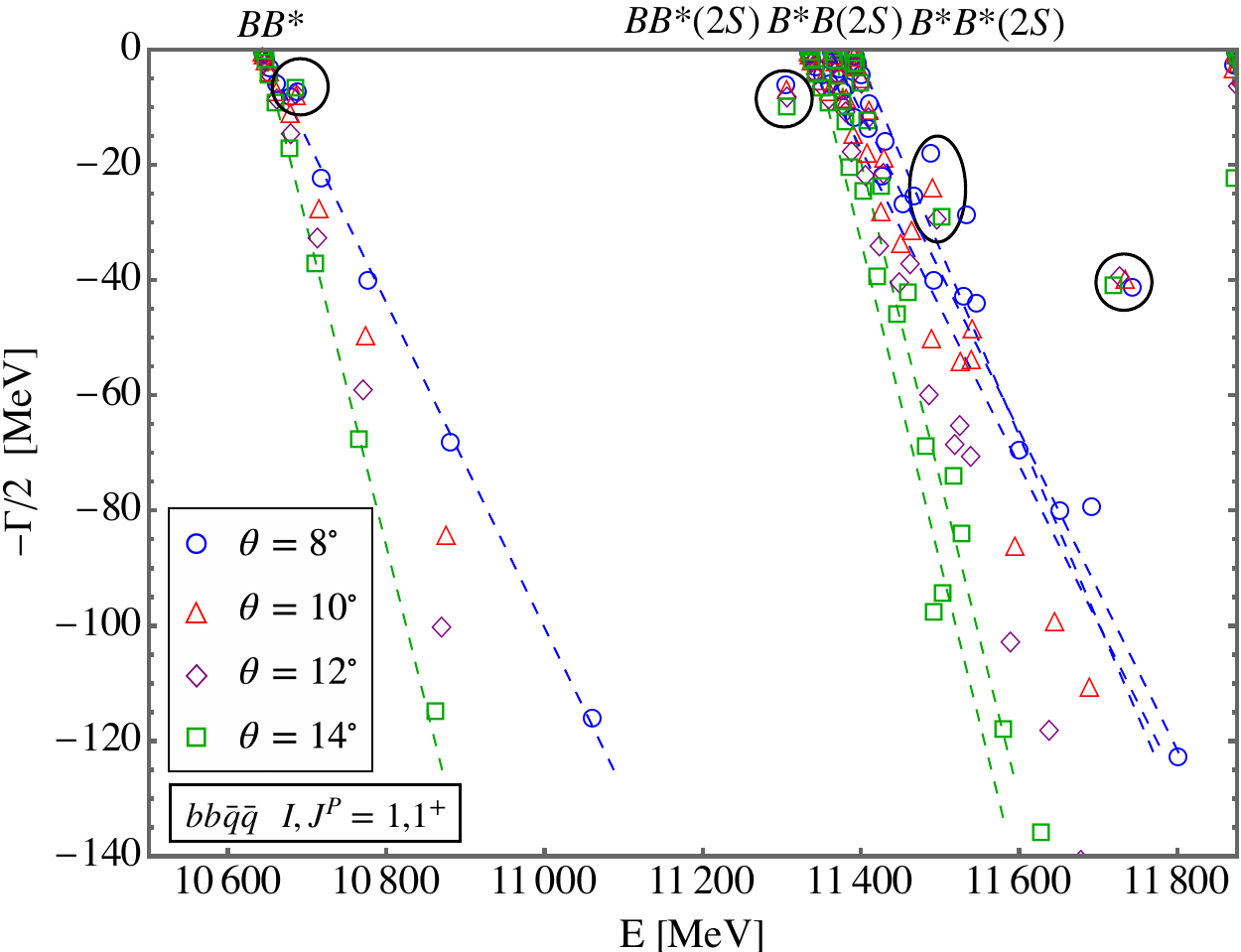}     \label{fig:bbqq1}

\end{minipage}%
}%
\subfigure[]{
\begin{minipage}[t]{0.35\linewidth}
\centering
\includegraphics[width=1\textwidth]{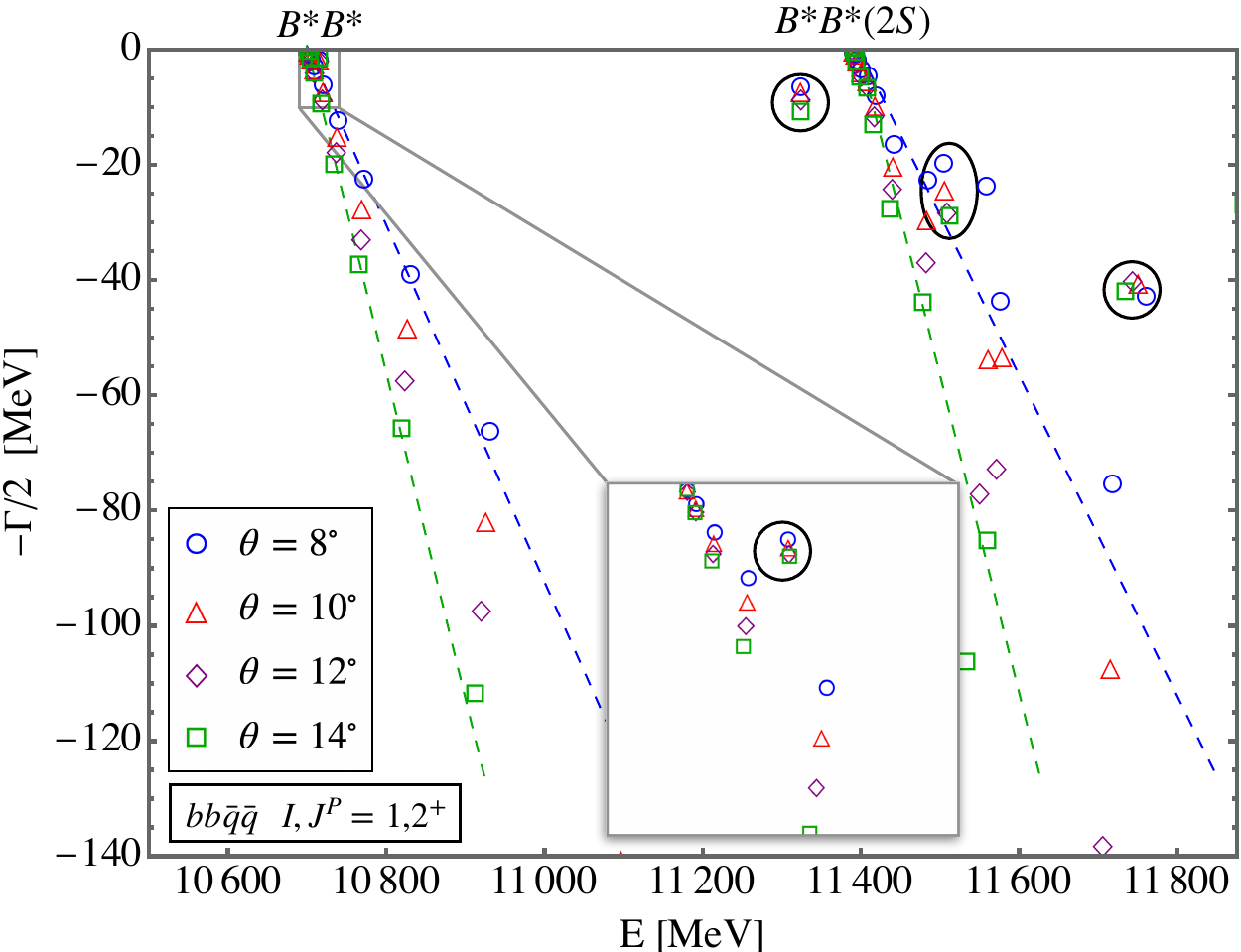}
    \label{fig:bbqq2}
\end{minipage}%

}%
\caption{ The complex eigenvalues of the $bb\bar{q}\bar{q}$ ($\bar b\bar b qq$) states with $I(J^{P})$ quantum numbers (a) $1(0^{+}$), (b) $1(1^{+})$, and (c) $1(2^{+})$. The results are obtained with varying $\theta$ in the complex-scaling method. The inset provides a detailed view of the selected energy region, highlighting the resonances marked in black circles.} \label{fig:bbqq}
\end{figure*}

\begin{figure*}
\centering \subfigure[]{
\begin{minipage}[t]{0.35\linewidth}
\includegraphics[width=1\textwidth]{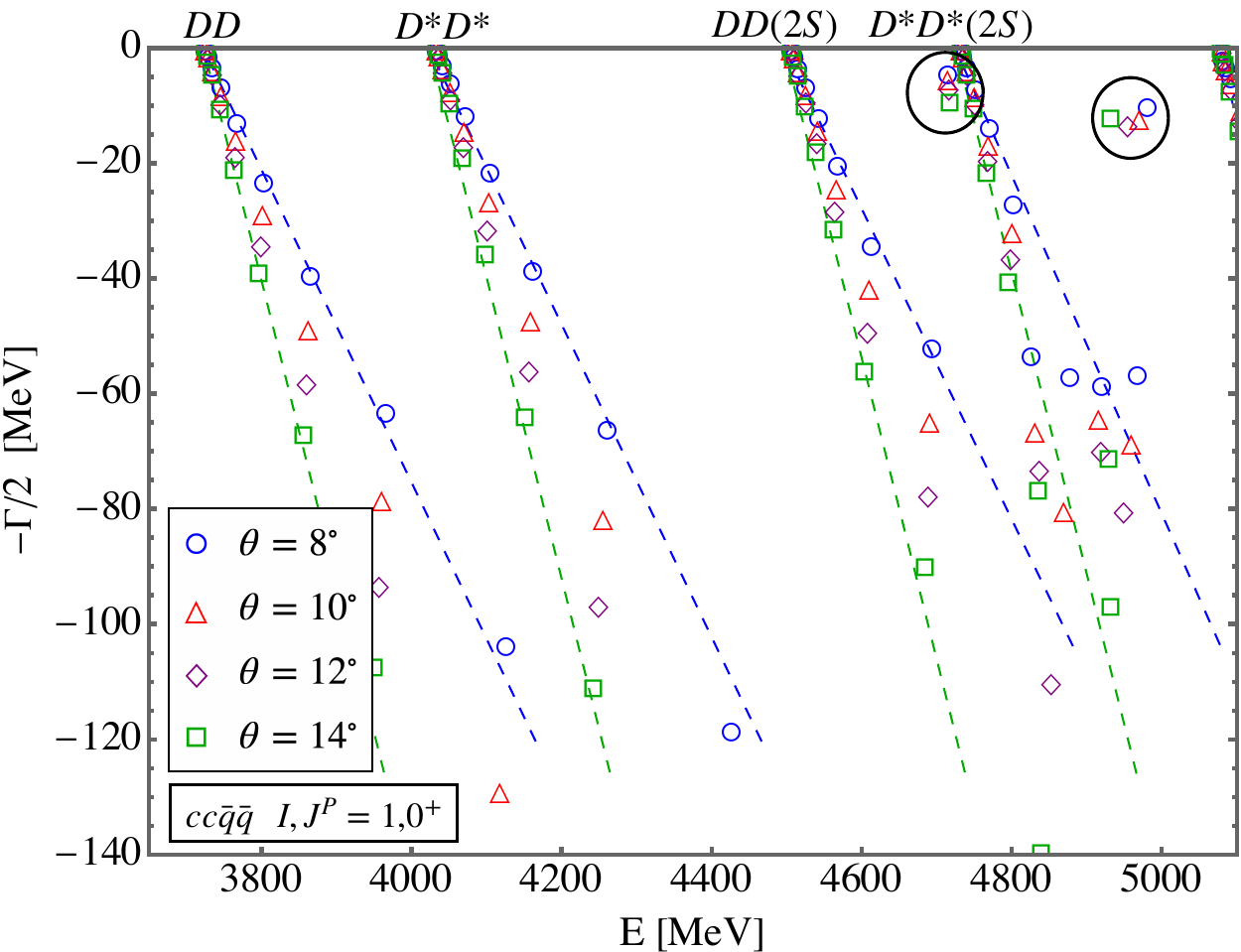}
 \label{fig:ccqq0}
\end{minipage}%
}%
\subfigure[]{
\begin{minipage}[t]{0.35\linewidth}
\centering
\includegraphics[width=1\textwidth]{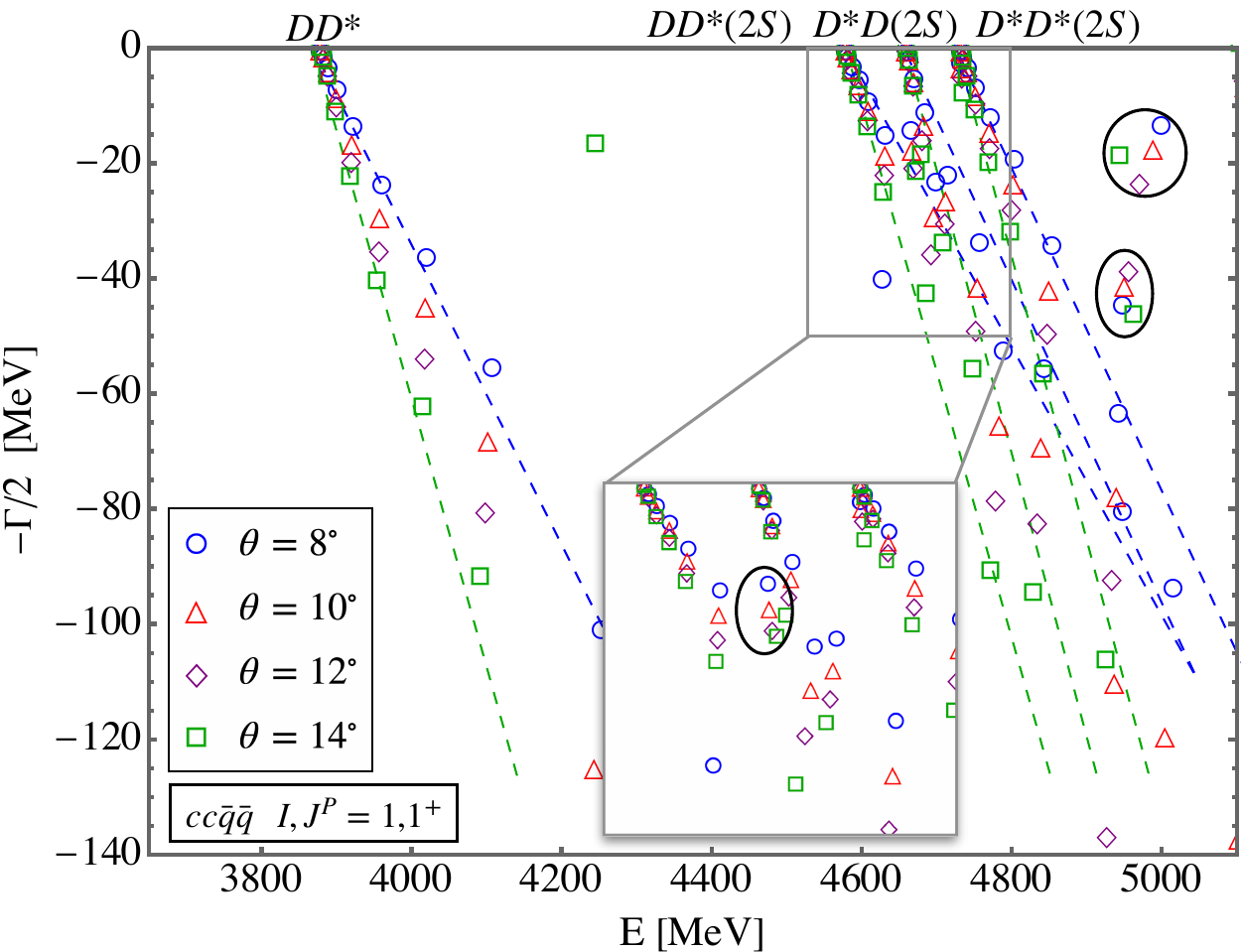}     \label{fig:ccqq1}

\end{minipage}%
}%
\subfigure[]{
\begin{minipage}[t]{0.35\linewidth}
\centering
\includegraphics[width=1\textwidth]{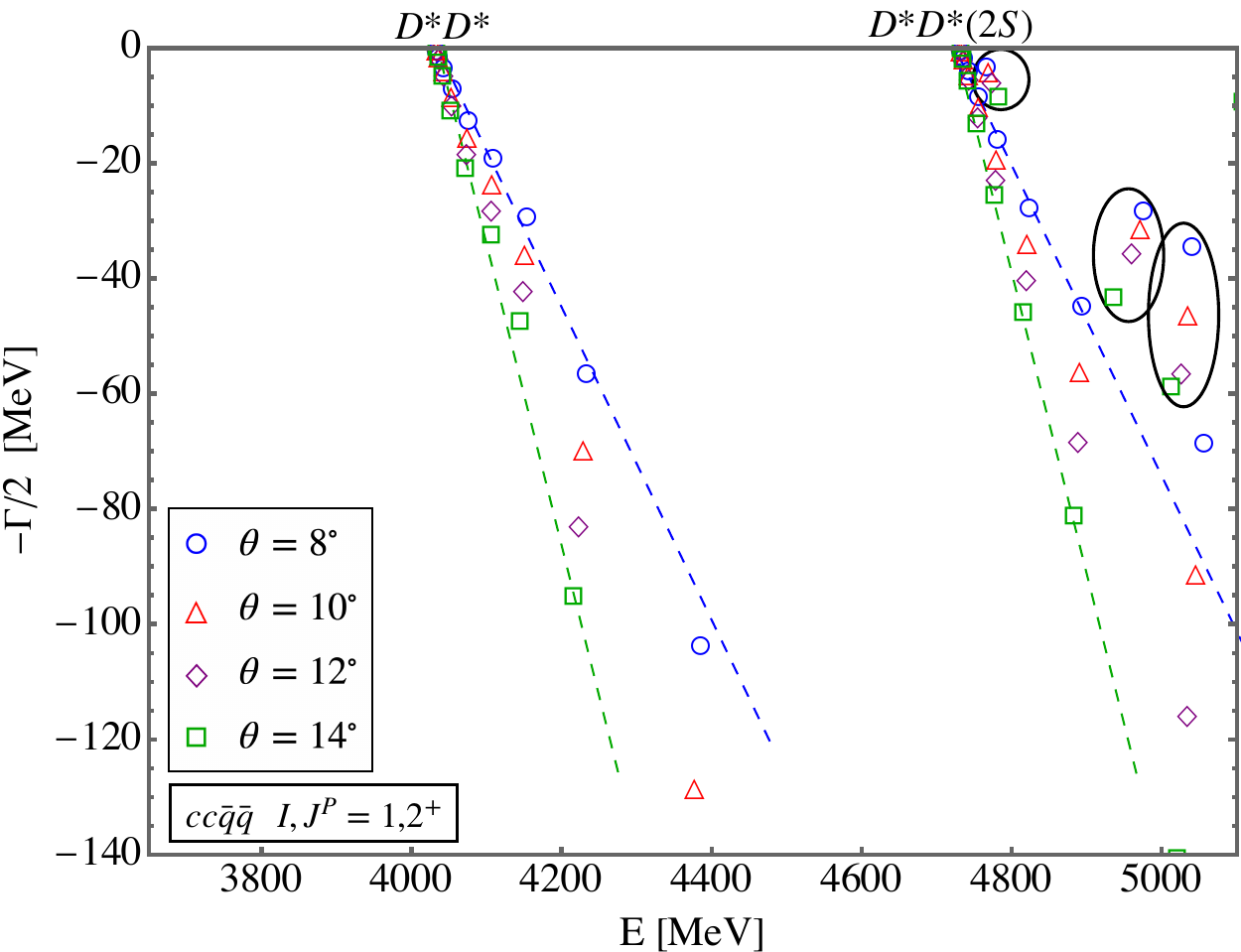}
    \label{fig:ccqq2}
\end{minipage}%

}%
\caption{ The complex eigenvalues of the $cc\bar{q}\bar{q}$ ($\bar c\bar c qq$) states with $I(J^{P})$ quantum numbers (a) $1(0^{+}$), (b) $1(1^{+})$, and (c) $1(2^{+})$. The results are obtained with varying $\theta$ in the complex-scaling method. The inset provides a detailed view of the selected energy region, highlighting the resonances marked in black circles.} \label{fig:ccqq}
\end{figure*}

\begin{figure*}
\centering \subfigure[]{
\begin{minipage}[t]{0.35\linewidth}
\includegraphics[width=1\textwidth]{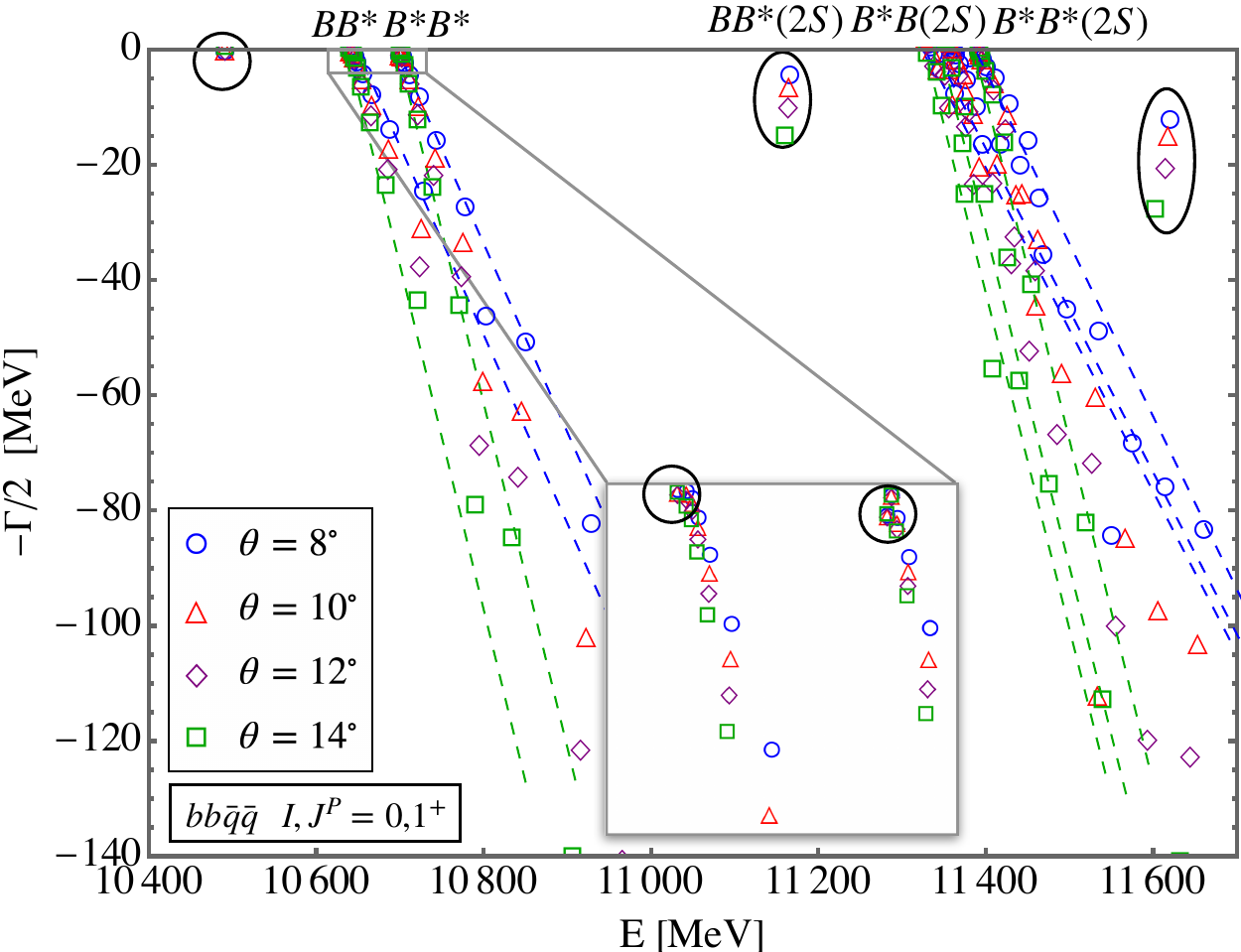}
 \label{fig:bbqq01}
\end{minipage}%
}%
\subfigure[]{
\begin{minipage}[t]{0.35\linewidth}
\centering
\includegraphics[width=1\textwidth]{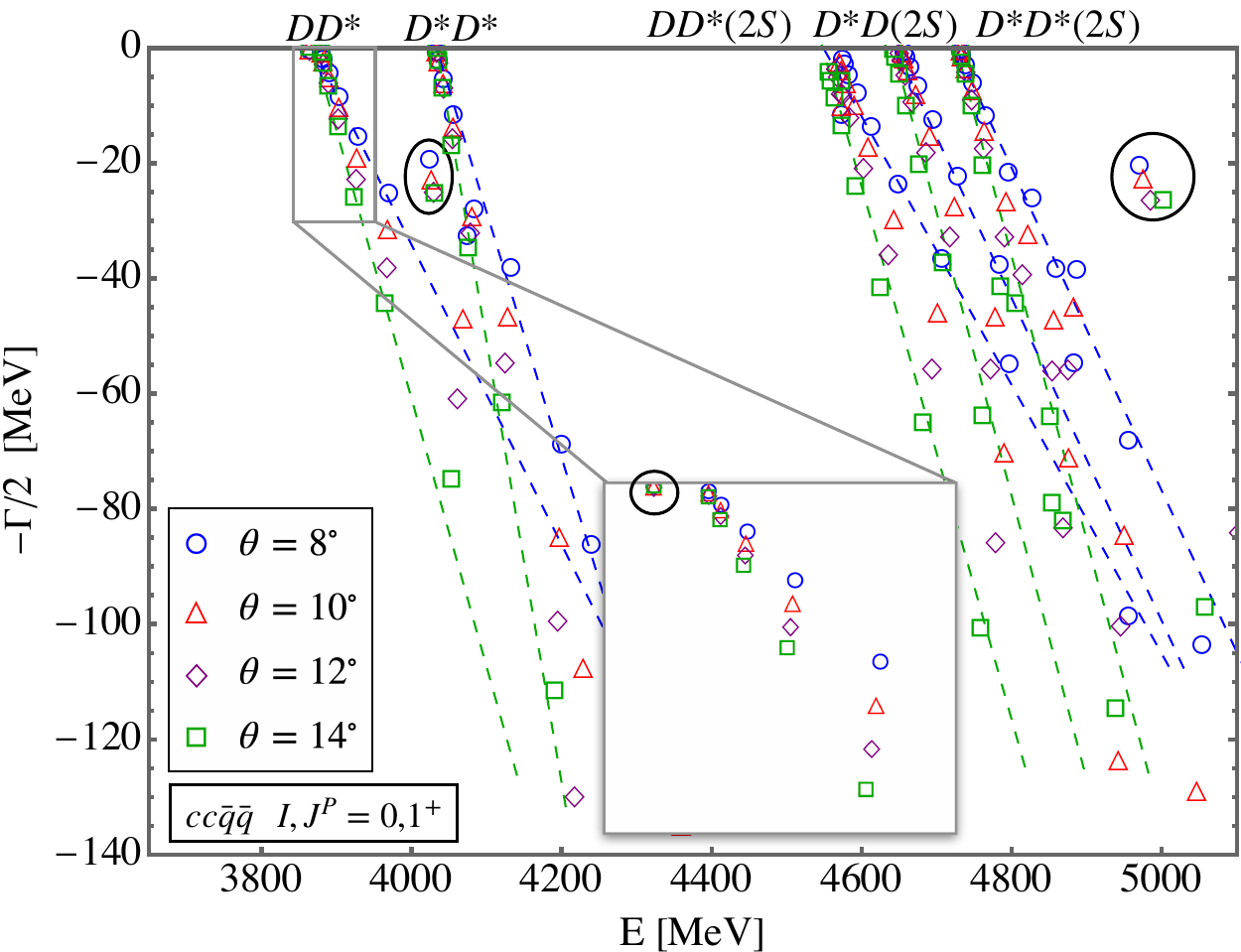}     \label{fig:ccqq01}

\end{minipage}%
}%

\caption{ The complex eigenvalues of the  (a) $bb\bar{q}\bar{q}$ ($\bar b\bar b qq$) states and (b) $cc\bar{q}\bar{q}$ ($\bar c\bar c qq$) states with  quantum number $I(J^{P})=0(1^{+})$. The results are obtained with varying $\theta$ in the complex-scaling method. The inset provides a detailed view of the selected energy region, highlighting the resonances marked in black circles.} \label{fig:iso0csm}
\end{figure*}

In this section, we list all the complex energy plots for the fully heavy tetraquark $QQ\bar Q'\bar Q'$ in Figs. \ref{fig:cccc}-\ref{fig:bbcc}, $QQ\bar s\bar s$ states in Figs. \ref{fig:bbss}-\ref{fig:ccss}, and $QQ\bar q\bar q$ states in Figs. \ref{fig:bbqq}-\ref{fig:iso0csm}, which have been used to identify the bound and resonant states. 

\end{appendix}

\bibliography{references.bib}

\end{document}